\documentclass[review]{elsarticle}

\usepackage[english]{babel} 
\usepackage[T1]{fontenc}
\usepackage{lmodern} 
\usepackage[ansinew]{inputenc}
\usepackage{amsmath}
\usepackage{amssymb}
\usepackage{xfrac}
\usepackage{tikz}
\usepackage{tikz-dimline}
\usepackage{lineno}

\pgfplotsset{
compat=1.5,
legend image code/.code={
\draw[mark repeat=2,mark phase=2]
plot coordinates {
(0cm,0cm)
(0.15cm,0cm)        
(0.3cm,0cm)         
};
}
}
\usepackage{geometry}
\usepackage{blindtext}
\usepackage{float}
\usepackage{graphicx}
\usepackage{xcolor}
\usepackage{caption}
\usepackage{hyperref}
\usepackage{pgfplots}
\usepackage{array}
\usepackage{siunitx}
\usepackage{multirow}
\usepackage{adjustbox} 
\usepackage{subcaption}
\usepackage{bm}

\usepackage{mathtools}
\usepackage[]{algorithm2e}
\usepackage{algpseudocode}
\usepackage[frozencache]{minted}
\usepgfplotslibrary{groupplots,dateplot}
\usetikzlibrary{patterns,shapes.arrows,calc,external,decorations,shapes,positioning}
\usepgfplotslibrary{fillbetween}
\tikzset{>=latex}
\pgfplotsset{compat=newest}

\begin{document}

\title{JAX-FLUIDS: A fully-differentiable high-order computational fluid dynamics solver for compressible two-phase flows}

\author{Deniz A. Bezgin\corref{cor}\fnref{fn}}
\ead{deniz.bezgin@tum.de}

\author{Aaron B. Buhendwa\corref{cor}\fnref{fn}}
\ead{aaron.buhendwa@tum.de}

\cortext[cor]{Corresponding author}
\fntext[fn]{Both authors contributed equally}

\author{Nikolaus A. Adams\corref{}}
\ead{nikolaus.adams@tum.de}

\address{Technical University of Munich, School of Engineering and Design, Chair of Aerodynamics and Fluid Mechanics, Boltzmannstr. 15, 85748 Garching bei M\"unchen, Germany}

\begin{frontmatter}
    \begin{abstract}
Physical systems are governed by partial differential equations (PDEs).
The Navier-Stokes equations describe fluid flows and are representative of nonlinear physical systems 
with complex spatio-temporal interactions.
Fluid flows are omnipresent in nature and engineering applications, and their accurate simulation is essential
for providing insights into these processes.
While PDEs are typically solved with numerical methods, the recent success of machine learning (ML)
has shown that ML methods can provide novel avenues of finding solutions to PDEs.
ML is becoming more and more present in computational fluid dynamics (CFD).
However, up to this date, there does not exist a general-purpose ML-CFD package which provides 1) powerful state-of-the-art 
numerical methods, 2) seamless hybridization of ML with CFD, and 3) automatic differentiation (AD) capabilities.
AD in particular is essential to ML-CFD research as it provides gradient information and enables 
optimization of preexisting and novel CFD models. 
In this work, we propose JAX-FLUIDS: a comprehensive fully-differentiable CFD Python solver for compressible 
two-phase flows.
JAX-FLUIDS allows the simulation of complex fluid dynamics with phenomena like three-dimensional turbulence, 
compressibility effects, and two-phase flows.
Written entirely in JAX, it is straightforward to include existing ML models into the proposed framework.
Furthermore, JAX-FLUIDS enables end-to-end optimization.
I.e., ML models can be optimized with gradients that are backpropagated through the entire CFD algorithm, and therefore
contain not only information of the underlying PDE but also of the applied numerical methods.
We believe that a Python package like JAX-FLUIDS is crucial to facilitate research at the intersection of ML and CFD
and may pave the way for an era of differentiable fluid dynamics.
\end{abstract}


\end{frontmatter}
\journal{}

\section{Introduction}
\label{sec:Intro}
The evolution of most known physical systems can be described by partial differential equations (PDEs).
Navier-Stokes equations (NSE) are partial differential equations that describe the continuum-scale flow of fluids.
Fluid flows are omnipresent in engineering applications and in nature,
and the accurate numerical simulation of complex flows is crucial for the prediction of global weather phenomena \cite{Bauer2015,Hafner2021}, 
for applications in biomedical engineering such as air flow through the lungs or blood circulation \cite{Nowak2003,Johnston2004},
and for the efficient design of turbomachinery \cite{Denton1998}, wind turbines \cite{Hansen2006,Sanderse2011},
and airplane wings \cite{Lyu2014}.
Computational fluid dynamics (CFD) aims to solve these problems with numerical algorithms.

While classical CFD has a rich history, in recent years the symbiosis of machine learning (ML) and CFD has 
sparked a great interest amongst researchers \cite{Duraisamy2019,Brenner2019,Brunton2020a}.
The amalgamation of classical CFD and ML requires powerful novel algorithms which allow seamless integration of data-driven models 
and, more importantly, end-to-end automatic differentiation \cite{Baydin2018} through the entire algorithm.
Here, we provide JAX-FLUIDS (\url{https://github.com/tumaer/JAXFLUIDS}): the first state-of-the-art fully-differentiable CFD framework for the computation of 
three-dimensional compressible two-phase flows with high-order numerical methods.
Over the course of this paper, we discuss the challenges of hybrid ML-accelerated CFD solvers and highlight 
how novel architectures like JAX-FLUIDS have the potential to facilitate ML-supported fluid dynamics research.

The quest for powerful numerical fluid dynamics algorithms has been a long-lasting challenge. 
In the mid of the last century, the development of performant computer processing units (CPUs) laid the fundament for the development 
of computational fluid dynamics (CFD).
For the first time, computers were used to simulate fluid flows \cite{Harlow2004}.
Computational fluid dynamics, i.e., the numerical investigation of fluid flows, became a scientific field on its own.  
With the rapid development of computational hardware, the CFD community witnessed new and powerful algorithms 
for the computation of more and more complex flows.
Among others, robust time integration schemes, high-order spatial discretizations, 
and accurate flux-functions were thoroughly investigated.
In the 1980s and 1990s, many advancements in numerical methods for compressible fluid flows followed, 
e.g., \cite{VanLeer1979,Roe1981,Woodward1984,Toro1994,Liou1996}.

In recent years, machine learning (ML) has invigorated the physical sciences by providing novel tools for
predicting the evolution of physical systems. 
Breakthrough inventions in ML \cite{Lecun1989,Hochreiter1997} and rapid technical developments of graphics processing units (GPUs) 
have led to an ever-growing interest in machine learning.
Starting from applications in computer vision \cite{LeCun2010,Lecun2015}, cognitive sciences \cite{Lake2015}, 
and genomics \cite{Alipanahi2015}, 
machine learning and data-driven methods have also become more and more popular in physical and engineering sciences.
This development has partially been fuelled by the emergence of powerful general-purpose automatic differentiation frameworks, 
such as Tensorflow \cite{Abadi}, Pytorch \cite{Paszke2019}, and JAX \cite{jax2018github}. 
Natural and engineering sciences can profit from ML methods as they have capabilities to learn complex relations from data and 
enable novel strategies for modelling physics as well as new avenues of post-processing.
For example, machine learning has been successfully used to identify PDEs from data \cite{Brunton2016}, 
and physics-informed neural networks provide new ways for solving inverse problems \cite{Raissi2019,Buhendwa2021c} 
by combining data and PDE knowledge.

It is well-known that fluid dynamics is a data-rich and compute-intensive discipline \cite{Brunton2020a}.
The nonlinearity of the underlying governing equations, the Navier-Stokes equations for viscous flows and the Euler equations 
for inviscid flows, is responsible for very complex spatio-temporal features of fluid flows.
For example, turbulent flows exhibit chaotic behavior with strong intermittent flow features, and non-Gaussian statistics.
At the same time, in the inviscid Euler equations strong discontinuities can form over time due to the hyperbolic character 
of the underlying PDEs.

Machine learning offers novel data-driven methods to tackle long-standing problems in fluid dynamics \cite{Duraisamy2019,Brunton2020a,Brunton2020b}.
A multitude of research has put forward different ways of incorporating ML in CFD applications.
Applications range from fully data-driven surrogate models to less invasive hybrid data-driven numerical methods.
Thereby, scientific ML methods can be categorized according to different criteria.

One important distinction is the level of physical prior-knowledge that is included in model and training \cite{Karniadakis2021}.
Entirely data-driven ML models have the advantage of being quickly implemented and efficient during inference.
However, they typically do not offer guarantees on performance (e.g., convergence, stability, and generalization),
convergence in training is challenging,
and it is often difficult to enforce physical constraints such as symmetries or conservation of energy.
In contrast, established numerical methods are consistent and enforce physical constraints.
Recently, intense research has investigated the hybridization of ML and classical numerical methods.

A second major distinction of ML models can be made according to on- and offline training.
Up until now, ML models have been typically optimized offline, i.e., outside of physics simulators.
Upon proper training, they are then plugged into an existing CFD solver for evaluation of down-stream tasks.
Examples include training of explicit subgrid scale models in large eddy simulations \cite{Beck2019}, 
interface reconstruction in multiphase flows \cite{Patel2019a,Buhendwa2021b}, 
and cell face reconstruction in shock-capturing schemes \cite{Stevens2020c,Bezgin2021b}.

Although the offline training of ML models is relatively easy, there are several drawbacks to this approach.
For one, these models suffer from a data-distribution mismatch between the data seen at training and test time.
Secondly, they generally do not directly profit from a priori knowledge about the dynamics of the underlying PDE.
Additionally, fluid mechanics solvers are often very complex, written in low-level programming languages like Fortran or C++, 
and heavily optimized for CPU computations.
This is in contrast with practices in ML research: ML models are typically trained in Python and optimized for GPU usage. 
Inserting these models into preexisting CFD software frameworks can be a tedious task.

To tackle this problem, researchers have come up with differentiable CFD frameworks written entirely in Python 
which allow end-to-end optimization of ML models.
The end-to-end (online) training approach utilizes automatic differentiation \cite{Baydin2018} through entire simulation trajectories.
ML models trained in such a fashion experience the PDE dynamics and also see their own outputs during training.
Among others, Bar-Sinai and co-workers \cite{Bar-Sinai2019,Zhuang2021} have proposed a differentiable framework for finding 
optimal discretizations for simple non-linear one-dimensional problems and turbulent mixing in two-dimensions.
In \cite{Um2020}, a differentiable solver was placed in the training loop to reduce the error of iterative solvers.
Bezgin et al. \cite{Bezgin2021a} have put forward a subgrid scale model for nonclassical shocks.
Kochkov et al. \cite{Kochkove2101784118} have optimized a subgrid scale model for two-dimensional turbulence.

Afore-noted works focus on simpler flow configurations. 
The problems are often one- and two-dimensional, incompressible, and lack complex physics such as two-phase flows, 
three-dimensional turbulence, or compressibility effects. 
Additionally, the written code packages often are highly problem specific and cannot be used as general differentiable CFD 
software packages.
To the knowledge of the authors, despite a high-interest in ML-accelerated CFD, to this day there does not exist a 
comprehensive mesh-based software package for \textit{differentiable fluid dynamics} in the Eulerian reference frame.
At the same time, for Lagrangian frameworks, the differentiable software package JAX-MD \cite{Schoenholz} has been successfully used 
for molecular dynamics and provides a general fundament for other particle-based discretization methods.

We reiterate that the steady rise and success of machine learning in computational fluid dynamics, 
but also more broadly in computational physics, calls for a new generation of algorithms which allow 
\begin{enumerate}
    \item rapid prototyping in high-level programming languages,
    \item algorithms which can be run on CPUs, GPUs, and TPUs,
    \item the seamless integration of machine learning models into solver frameworks,
    \item fully-differentiable algorithms which allow end-to-end optimization of data-driven models.
\end{enumerate}

Realizing the advent of \textit{differentiable fluid dynamics} and the increasing need for a differentiable general high-order 
CFD solver, here we introduce JAX-FLUIDS as a fully-differentiable general-purpose 3D finite-volume CFD 
solver for compressible two-phase flows. 
JAX-FLUIDS is written entirely in JAX \cite{jax2018github}, a numerical computing library with automatic differentiation capabilities 
which has seen increased popularity in the physical science community over the past several years. 
JAX-FLUIDS provides a wide variety of state-of-the-art high-order methods for compressible turbulent flows.
A powerful level-set implementation allows the simulation of arbitrary solid boundaries and two-phase flows.
High-order shock-capturing methods enable the accurate computation of shock-dominated compressible flow problems and 
shock-turbulence interactions.
Performance and usability were key design goals during development of JAX-FLUIDS.
The solver can conveniently be installed as a Python package.
The source code builds on the JAX NumPy API. Therefore, JAX-FLUIDS is accessible, performant, and runs on accelerators like CPUs, GPUs, and TPUs.
Additionally, an object-oriented programming style and a modular design philosophy allows users the easy integration of new modules.
We believe that frameworks like JAX-FLUIDS are crucial to facilitate research at the intersection of ML and CFD,
and may pave the way for an era of \textit{differentiable fluid dynamics}.

The remainder of this paper is organized as follows.
In Sections \ref{sec:PhysicalModel} and \ref{sec:numericalmodel}, we describe the physical and numerical model. 
Section \ref{sec:Implementation} describes challenges of writing a high-performance CFD code in a high-level programming language 
such as JAX.
We additionally detail the general structure of our research code.
In Section \ref{sec:ForwardPass}, we evaluate our code as a classical physics simulator. 
We validate the numerical framework on canonical test cases from compressible fluid mechanics, including strong-shocks and turbulent flows.
In Section \ref{sec:Performance}, we assess the computational performance of JAX-FLUIDS.
Section \ref{sec:BackwardPass} showcases how the proposed framework can be used in machine learning tasks.
Specifically, we demonstrate full-differentiability of the framework by optimizing a numerical flux function.
Finally, in Section \ref{sec:Conclusion} we conclude and summarize the main results of our work.
\section{Physical Model}
\label{sec:PhysicalModel}
We are interested in the compressible Euler equations for inviscid flows and in the compressible Navier-Stokes equations (NSE)
which govern viscous flows.
The state of a fluid at any position in the flow field $\mathbf{x} = \left[ x, y, z \right]^T = \left[ x_1, x_2, x_3 \right]^T$ at time $t$ can be described by the vector of primitive variables 
$\mathbf{W} = \left[ \rho, u, v, w, p \right]^T$.
Here, $\rho$ is the density, $\mathbf{u} = \left[ u, v, w \right]^T = \left[ u_1, u_2, u_3 \right]^T$ is the velocity vector, and $p$ is the pressure.
An alternative description is given by the vector of conservative variables 
$\mathbf{U} = \left[ \rho, \rho u, \rho v, \rho w, E \right]^T$.
Here, $\rho \mathbf{u} = \left[ \rho u, \rho v, \rho w \right]^T$ are the momenta in the three spatial dimensions,
and $E = \rho e + \frac{1}{2} \rho \mathbf{u} \cdot \mathbf{u} $ is the total energy per unit volume.
$e$ is the internal energy per unit mass.

In differential formulation, the compressible Euler equations can be written in terms of $\mathbf{U}$
\begin{equation}
    \frac{\partial \mathbf{U}}{\partial t} + \frac{\partial \mathcal{F}(\mathbf{U})}{\partial x} + \frac{\partial \mathcal{G}(\mathbf{U})}{\partial y} + \frac{\partial \mathcal{H}(\mathbf{U})}{\partial z} = 0.
    \label{eq:DiffConsLaw1}
\end{equation}
The convective physical fluxes $\mathcal{F}$, $\mathcal{G}$, and $\mathcal{H}$ are defined as
\begin{align}
    \mathcal{F}(\mathbf{U}) = \begin{pmatrix}
        \rho u \\
        \rho u^2 + p \\
        \rho u v  \\
        \rho u w \\
        u (E + p)
    \end{pmatrix}, \quad
    \mathcal{G}(\mathbf{U}) = \begin{pmatrix}
        \rho v \\
        \rho v u\\
        \rho v^2 + p \\
        \rho v w \\
        v (E + p)
    \end{pmatrix}, \quad
    \mathcal{H}(\mathbf{U}) = \begin{pmatrix}
        \rho w\\
        \rho w u\\
        \rho w v \\
        \rho w^2 + p \\
        w (E + p)
    \end{pmatrix}.
\end{align}
This set of equations must be closed by an equation of state (EOS) which relates pressure with density and internal energy, 
i.e., $p = p(\rho, e)$.    
Unless specified otherwise, we use the stiffened gas equation 
\begin{equation}
    p(\rho, e) = (\gamma - 1) \rho e - \gamma B,
    \label{eq:StiffenedGas}
\end{equation}
where $\gamma$ represents the ratio of specific heats and $B$ is the background pressure.

The compressible Navier-Stokes equations can be seen as the viscous extension of the Euler equations.
As before, we write them in terms of the conservative state vector $\mathbf{U}$,
\begin{align}
    \frac{\partial \mathbf{U}}{\partial t} + \frac{\partial \mathcal{F}(\mathbf{U})}{\partial x} + \frac{\partial \mathcal{G}(\mathbf{U})}{\partial y} + \frac{\partial \mathcal{H}(\mathbf{U})}{\partial z} = 
    \frac{\partial \mathcal{F}^{d}(\mathbf{U})}{\partial x} + \frac{\partial \mathcal{G}^{d}(\mathbf{U})}{\partial y} + \frac{\partial \mathcal{H}^{d}(\mathbf{U})}{\partial z} + S(\mathbf{U}).
    \label{eq:DiffConsLaw2}
\end{align}
Here, we have additionally introduced the dissipative fluxes $\mathcal{F}^d$, $\mathcal{G}^d$, and $\mathcal{H}^d$ and 
the source term vector $S(\mathbf{U})$ on the right-hand side.
The dissipative fluxes describe viscous effects and heat conduction, and are given by
\begin{align}
    \mathcal{F}^d(\mathbf{U}) = \begin{pmatrix}
        0 \\
        \tau^{11} \\
        \tau^{12} \\
        \tau^{13} \\
        \sum_i u_i \tau^{1i} - q_1
    \end{pmatrix}, \quad
    \mathcal{G}^d(\mathbf{U}) = \begin{pmatrix}
        0 \\
        \tau^{21} \\
        \tau^{22} \\
        \tau^{23} \\
        \sum_i u_i \tau^{2i} - q_2
    \end{pmatrix}, \quad
    \mathcal{H}^d(\mathbf{U}) = \begin{pmatrix}
        0 \\
        \tau^{31} \\
        \tau^{32} \\
        \tau^{33} \\
        \sum_i u_i \tau^{3i} - q_3
    \end{pmatrix}.
\end{align}
The stresses $\tau^{ij}$ are given by
\begin{align}
    \tau^{ij} = \mu \left(\frac{\partial u_i}{\partial x_j} + \frac{\partial u_j}{\partial x_i}\right) - \frac{2}{3} \mu \delta_{ij} \frac{\partial u_k}{\partial x_k}.
\end{align}
$\mu$ is the dynamic viscosity.
The energy flux vector $\mathbf{q}$ can be expressed via Fourier's heat conduction law, $\mathbf{q} = [q_1, q_2, q_3]^T = - \lambda \nabla T$.
$\lambda$ is the heat conductivity.
The source terms $S(\mathbf{U})$ represent body forces or heat sources.
The body force resulting from gravitational acceleration $\mathbf{g}=[g_1, g_2, g_3]^T$ 
is given by
\begin{equation}
    S(\mathbf{U}) =
    \begin{pmatrix}
        0 \\
        \rho\mathbf{g} \\
        \rho \mathbf{u} \cdot \mathbf{g}
    \end{pmatrix}.
\end{equation}

In Table \ref{tab:Nondimensionalization}, we summarize the nomenclature and the reference values with which we non-dimensionalize aforementioned equations.
\begin{table}[t]
    \begin{center}
        \small
        \begin{tabular}{ l  l  l  } 
        \hline
        Quantitiy & Nomenclature & Reference quantity \\
        \hline
        Density                     & $\rho$            & $\rho_{ref}$      \\ 
        Length                      & $(x, y, z)$ or $(x_1, x_2, x_3)$         & $l_{ref}$         \\ 
        Velocity                    & $(u, v, w)$ or $(u_1, u_2, u_3)$   & $u_{ref}$         \\ 
        Temperature                 & $T$               & $T_{ref}$         \\
        Time                        & $t$               & $t_{ref} = l_{ref} / u_{ref}$     \\
        Pressure                    & $p$               & $p_{ref} = \rho_{ref} u_{ref}^2$  \\
        Viscosity                   & $\mu$             & $\mu_{ref} = \rho_{ref} u_{ref} l_{ref}$ \\
        Surface tension coefficient & $\sigma$          & $\sigma_{ref} = \rho_{ref} u_{ref}^2 l_{ref}$ \\
        Thermal conductivity        & $\lambda$          & $\lambda_{ref} = \rho_{ref} u_{ref}^3 l_{ref} / T_{ref}$ \\
        Gravitation                 & $g$               & $g_{ref} = u_{ref}^2 / l_{ref}$ \\
        Specific gas constant       & $\mathcal{R}$     & $\mathcal{R}_{ref} = u_{ref}^2 / T_{ref}$ \\
        Mass                        & $m$               & $m_{ref} = \rho_{ref} l_{ref}^3$ \\    
        Mass flow                   & $\dot{m}$         & $\dot{m}_{ref} = m_{ref} / t_{ref} = \rho_{ref} u_{ref} l_{ref}^2$ \\    
        \hline
    \end{tabular}
    \caption{Overview on nomenclature and nondimensionalization.}
    \label{tab:Nondimensionalization}
    \end{center}
\end{table}
\section{Numerical Model}
\label{sec:numericalmodel}

In this section we detail the numerical methods of JAX-FLUIDS.
Table \ref{tab:NumericalMethods} provides an overview on the implemented numerical methods.

\subsection{Finite-Volume Discretization}
\label{subsec:FVD}
The differential form of the conservation law in Equations \eqref{eq:DiffConsLaw1} and \eqref{eq:DiffConsLaw2} assume smooth solutions 
for which partial derivatives exist.
In practice, we solve the integral form of the partial differential equations using the finite-volume method.
We use cuboid cells on a Cartesian grid.
In general, cell $(i,j,k)$ has spatial extension $\Delta x$, $\Delta y$, and $\Delta z$ in the spatial dimensions $x$, $y$, and $z$, 
respectively. We denote the corresponding cell volume as $V=\Delta x\Delta y\Delta z$.
Often, we use cubic cells for which $\Delta x = \Delta y = \Delta z$. 

In the finite-volume formulation we are interested in the spatio-temporal distribution of cell-averaged values which are defined by
\begin{equation}
    \bar{\mathbf{U}}_{i,j,k} = \frac{1}{V} \int_{x_{i-\frac{1}{2},j,k}}^{x_{i+\frac{1}{2},j,k}} \int_{y_{i,j-\frac{1}{2},k}}^{y_{i,j+\frac{1}{2},k}} \int_{z_{i,j,k-\frac{1}{2}}}^{z_{i,j,k+\frac{1}{2}}} \mathbf{U} \text{d}x\text{d}y\text{d}z. 
\end{equation}
After application of volume integration to Equation \eqref{eq:DiffConsLaw2}, the temporal evolution of the cell-averaged value in cell $(i,j,k)$ is given by 
\begin{align}
    \begin{split}
        \frac{\text{d}}{\text{d}t} \bar{\mathbf{U}}_{i,j,k} = &- \frac{1}{\Delta x} \left(\mathbf{F}_{i+\frac{1}{2},j,k} - \mathbf{F}_{i-\frac{1}{2},j,k} \right) \\
        & - \frac{1}{\Delta y} \left( \mathbf{G}_{i,j+\frac{1}{2},k} - \mathbf{G}_{i,j-\frac{1}{2},k} \right) \\
        & - \frac{1}{\Delta z} \left( \mathbf{H}_{i,j,k+\frac{1}{2}} - \mathbf{H}_{i,j,k-\frac{1}{2}}\right) \\
        & + \frac{1}{\Delta x} \left(\mathbf{F}^d_{i+\frac{1}{2},j,k} - \mathbf{F}^d_{i-\frac{1}{2},j,k} \right) \\
        & + \frac{1}{\Delta y} \left( \mathbf{G}^d_{i,j+\frac{1}{2},k} - \mathbf{G}^d_{i,j-\frac{1}{2},k} \right) \\
        & + \frac{1}{\Delta z} \left( \mathbf{H}^d_{i,j,k+\frac{1}{2}} - \mathbf{H}^d_{i,j,k-\frac{1}{2}}\right) \\
        & + \bar{\mathbf{S}}_{i,j,k}.
       \label{eq:FVD} 
    \end{split}
\end{align}
$\mathbf{F}$, $\mathbf{G}$, and $\mathbf{H}$ are the convective intercell numerical fluxes across the cell faces in $x$-, $y$-, and $z$-direction, and 
$\mathbf{F}^d$, $\mathbf{G}^d$, and $\mathbf{H}^d$ are the numerical approximations to the dissipative fluxes in $x$-, $y$-, and $z$-direction.
$\bar{\mathbf{S}}_{i,j,k}$ is the cell-averaged source term. 
Note that the convective and dissipative fluxes in Equation \eqref{eq:FVD} are cell face averaged quantities.
We use a simple one-point Gaussian quadrature to evaluate the necessary cell face integrals, however approximations of higher order are also possible, see for example \cite{Coralic2014e}.
For the calculation of the convective intercell numerical fluxes, we use WENO-type high-order discretization schemes \cite{Jiang1996} in combination with approximate Riemann solvers.
Subsection \ref{subsec:InviscidFlux} provides more details.
Dissipative fluxes are calculated by central finite-differences, see Subsection \ref{subsec:Dissipative} for details.
Multidimensional settings are treated via the dimension-by-dimension technique in a straightforward manner, i.e., aforementioned steps are repeated for each spatial dimension separately. 

\subsection{Time Integration}
\label{subsec:timeint}
The finite-volume discretization yields a set of ordinary differential equations (ODEs) \eqref{eq:FVD} that can be integrated in time by an ODE solver of choice.
We typically use explicit total-variation diminishing (TVD) Runge-Kutta methods \cite{Gottlieb1998a}. The time step size is given by the $CFL$ criterion 
as described in \cite{Hoppe2022a}.

\subsection{Convective Flux Calculation}
\label{subsec:InviscidFlux}
For the calculation of the convective fluxes an approximative solution to a Riemann problem has to be found at each cell face.
As we make use of the dimension-by-dimension technique, without loss of generality, we restrict ourselves to a one-dimensional setting
in this subsection.
We are interested in finding the cell face flux $\mathbf{F}_{i+\frac{1}{2}}$ at the cell face $x_{i+\frac{1}{2}}$.

We distinguish between two different methods for the calculation of the convective intercell fluxes: 
the \textit{High-order Godunov} approach and the \textit{Flux-splitting} approach.
In the \textit{High-order Godunov} approach, we first reconstruct flow states left and right of a cell face, and subsequently 
enter an approximate Riemann solver for the calculation of the convective intercell flux.
In the \textit{Flux-splitting} approach, we first perform cell-wise flux-splitting, reconstruct left and right going fluxes for each cell face,
and subsequently assemble the final flux.
In the following, we briefly sketch both methods.\\

\noindent\textit{High-order Godunov approach}
\begin{enumerate}
    \item Apply WENO reconstruction on the primitive variable $\mathbf{W}_i$/the conservative variable $\mathbf{U}_i$ to obtain the 
        cell face quantities $\mathbf{W}^{\pm}_{i+\frac{1}{2}}$/$\mathbf{U}^{\pm}_{i+\frac{1}{2}}$. 
        Note that the reconstruction can be done directly in physical space or via transformation in characteristic space.
        High-order reconstruction in physical space can lead to spurious oscillations due to the interaction of discontinuities
        in different fields \cite{Qiu2002}.
    \item Transform the reconstructed primitive/conservative variables at the cell face into conservative/primitive cell face quantities: 
    $\mathbf{W}^{\pm}_{i+\frac{1}{2}} \rightarrow \mathbf{U}^{\pm}_{i+\frac{1}{2}}$ / $\mathbf{U}^{\pm}_{i+\frac{1}{2}} \rightarrow \mathbf{W}^{\pm}_{i+\frac{1}{2}}$. 
    \item Compute the final flux with an appropriate flux function/approximate Riemann solver, e.g., HLL \cite{Harten1983a} or HLLC \cite{Toro1994}: 
    \begin{equation}
        \mathbf{F}_{i+\frac{1}{2}} = \mathbf{F}_{i+\frac{1}{2}} \left( \mathbf{U}_{i+\frac{1}{2}}^{-}, \mathbf{U}_{i+\frac{1}{2}}^{+}, \mathbf{W}_{i+\frac{1}{2}}^{-}, \mathbf{W}_{i+\frac{1}{2}}^{+} \right)
        \label{eq:ApproxRiemannSolver}
    \end{equation}
\end{enumerate}

\noindent\textit{Flux-Splitting approach}
\begin{enumerate}
    \item At the cell face $x_{i+\frac{1}{2}}$ compute an appropriate average state from neighboring cells $\mathbf{U}_{i+\frac{1}{2}} = \mathbf{U}_{i+\frac{1}{2}} \left( \mathbf{U}_{i}, \mathbf{U}_{i+1} \right)$ (e.g., by arithmetic mean or Roe average \cite{Toro2009a})
        and the corresponding Jacobian $\mathbf{A}_{i+\frac{1}{2}} = \mathbf{A}_{i+\frac{1}{2}} \left( \mathbf{U}_{i+\frac{1}{2}} \right)$.
    \item Eigenvalue decomposition of the Jacobian: $\mathbf{A}_{i+\frac{1}{2}} = \mathbf{R}_{i+\frac{1}{2}} \mathbf{\Lambda}_{i+\frac{1}{2}} \mathbf{R}_{i+\frac{1}{2}}^{-1}$, with the matrix of right eigenvectors $\mathbf{R}_{i+\frac{1}{2}}$,
        the matrix of left eigenvectors $\mathbf{R}_{i+\frac{1}{2}}^{-1}$, and the eigenvalues $\mathbf{\Lambda}_{i+\frac{1}{2}}$.
    \item Transform the cell state $\mathbf{U}_i$ and the flux $\mathbf{F}_i$ to characteristic space: $\mathbf{V}_j = \mathbf{R}_{i+\frac{1}{2}}^{-1} \mathbf{U}_i, \quad \mathbf{G}_j = \mathbf{R}_{i+\frac{1}{2}}^{-1} \mathbf{F}_i$.
    \item Perform the user-specified flux splitting: $\hat{\mathbf{G}}^{\pm}_i = \frac{1}{2} \left( \mathbf{G}_i \pm \bar{\mathbf{\Lambda}}_{i+\frac{1}{2}} \mathbf{V}_i \right)$, where $\bar{\mathbf{\Lambda}}_{i+\frac{1}{2}}$ is the eigenvalue matrix of the respective flux-splitting scheme.
    \item Apply WENO reconstruction on $\hat{\mathbf{G}}^{\pm}_i$ to obtain $\hat{\mathbf{G}}^{\pm}_{i+\frac{1}{2}}$ at the cell face $x_{i+\frac{1}{2}}$.
    \item Assemble the final flux in characteristic space: $\hat{\mathbf{G}}_{i+\frac{1}{2}} = \hat{\mathbf{G}}^{+}_{i+\frac{1}{2}} + \hat{\mathbf{G}}^{-}_{i+\frac{1}{2}}$.
    \item Transform the final flux back to physical space: $\mathbf{F}_{i+\frac{1}{2}} = \mathbf{R}_{i+\frac{1}{2}} \hat{\mathbf{G}}_{i+\frac{1}{2}}$.
\end{enumerate}

\subsection{Dissipative Flux Calculation}
\label{subsec:Dissipative}
For the calculation of the dissipative fluxes, we have to evaluate derivatives at the cell faces.
We do this using central finite-differences.
As we do all computations in a Cartesian framework, central finite-differences can be evaluated directly at a cell face if the direction of the derivative is 
parallel to the cell face normal.
For example, at the cell face $x_{i+1/2,j,k}$ the $x$-derivative of any quantity $\psi$ is directly approximated with second-order or fourth-order
central finite-differences, 
\begin{equation}
    \left.\frac{\partial \psi}{\partial x}\right\vert_{x_{i+1/2,j,k}}^{C2} = \frac{-\psi_{i,j,k} + \psi_{i+1,j,k}}{\Delta x}, \quad \left.\frac{\partial \psi}{\partial x}\right\vert_{x_{i+1/2,j,k}}^{C4} = \frac{\psi_{i-1,j,k} - 27 \psi_{i,j,k} + 27 \psi_{i+1,j,k} - \psi_{i+2,j,k}}{24 \Delta x}.
\end{equation}
If the cell face normal is perpendicular to the direction of the derivative, we use a two-step process to approximate the derivative.
We first evaluate the derivative at the cell-centers and then use a central interpolation to obtain the value at the cell face of interest.
Again, we use second-order or fourth-order order central finite-differences for the derivatives.
Let us consider the $y$-derivative of the quantity $\psi$ at the cell face $x_{i+\frac{1}{2},j,k}$ (the $z$-derivative is complelety analogous).
We first calculate the derivative at the cell centers
\begin{equation}
    \left.\frac{\partial \psi}{\partial y}\right\vert_{x_{i,j,k}}^{C2} = \frac{-\psi_{i,j-1,k} + \psi_{i,j+1,k}}{\Delta y}, \quad \left.\frac{\partial \psi}{\partial y}\right\vert_{x_{i,j,k}}^{C4} = \frac{\psi_{i,j-2,k} - 8 \psi_{i,j-1,k} + 8 \psi_{i,j+1,k} - \psi_{i,j+2,k}}{12 \Delta y}.
\end{equation}
We subsequently interpolate these values to the cell face,
\begin{align}
\begin{split}
    &\left.\frac{\partial \psi}{\partial y}\right\vert_{x_{i+1/2,j,k}} = \frac{1}{2} \left( \left.\frac{\partial \psi}{\partial y}\right\vert_{x_{i-1,j,k}} + \left.\frac{\partial \psi}{\partial y}\right\vert_{x_{i+1,j,k}} \right),\\ 
    &\left.\frac{\partial \psi}{\partial y}\right\vert_{x_{i+1/2,j,k}} = \frac{1}{16} \left( - \left.\frac{\partial \psi}{\partial y}\right\vert_{x_{i-2,j,k}} + 9 \left.\frac{\partial \psi}{\partial y}\right\vert_{x_{i-1,j,k}} + 9 \left.\frac{\partial \psi}{\partial y}\right\vert_{x_{i+1,j,k}} - \left.\frac{\partial \psi}{\partial y}\right\vert_{x_{i+2,j,k}} \right).
\end{split}
\end{align}

\subsection{Source Terms and Forcings}
\label{subsec:forcing}
The source terms $S(\mathbf{U})$ represent body forces and heat sources. We use them to impose physical constraints, e.g.,
fixed mass flow rates or temperature profiles. These forcings are required to simulate a variety of test cases. Examples include 
channel flows or driven homogenous isotropic turbulence.
Fixed mass flow rates are enforced with a PID controller minimizing the error
between target and current mass flow rate $e(t) = \frac{\dot{m}_{\text{target}} - \dot{m}(t)}{\dot{m}_{\text{target}}}$.
Here, the control variable is an acceleration $a_{\dot{m}}$ that drives the fluid in the prescribed direction.
We denote the unit vector pointing towards the prescribed direction as $\mathbf{N}$.
The controller variable and resulting source terms read
\begin{equation}
    a_{\dot{m}} = K_p e(t) + K_I \int_0^t e(\tau)\text{d}\tau + K_d \frac{\text{d}e(t)}{\text{d}t}, \qquad
    S(\mathbf{U}) =
    \begin{pmatrix}
        0 \\
        \rho a_{\dot{m}} \mathbf{N} \\
        \rho a_{\dot{m}} \mathbf{u} \cdot \mathbf{N}
    \end{pmatrix},
    \label{eq:PID_massflow}
\end{equation}
where $K_p$, $K_I$, and $K_d$ are the controller parameters. The integral and derivative in Equation \eqref{eq:PID_massflow} are 
approximated with first order schemes.
Fixed temperature profiles are enforced with a heat source $\dot{\omega}_T$. The heat source and resulting source term is given by
\begin{equation}
    \dot{\omega}_T =  \rho R \frac{\gamma}{\gamma - 1} \frac{T_{\text{target}} - T}{\Delta t}, \qquad S(\mathbf{U}) = \dot{\omega}_T [0,0,0,0,1]^T.
\end{equation}

\subsection{Level-set Method for Two-phase Flows}
\label{subsec:levelset}
\begin{figure}[t]
    \centering
    \begin{tikzpicture}

    \coordinate (NULL) at (0,0);
    \coordinate (B) at (5,5);
    \coordinate (C) at ($0.5*(B)$);
    \coordinate (D1) at ($0.125*(B)$);
    \coordinate (D2) at ($0.71*(B)$);

    \coordinate (D11) at ($0.1*(B)$);

    \coordinate (L) at ($1.1*(B|-NULL)+0.96*(B-|NULL)$);

    \filldraw[fill=black!10!white, line width=0.1pt] (NULL) -- (B|-NULL) -- ($(B|-NULL) + 0.16*(B-|NULL)$) -- ($(B|-NULL) + 0.16*(B-|NULL)$) arc (33.5:56.4:15) -- (B-|NULL) -- (NULL);

    \draw[line width=1pt] (0,0) rectangle (B);
    \draw[line width=0.5pt] (C |- NULL) -- (C |- B);
    \draw[line width=0.5pt] (C -| NULL) -- (C -| B);
    \draw[fill=red, draw=red] (C) circle (0.1);

    \draw[dashed] ($(C)+(C-|NULL)$) -- ($(C)+(B-|NULL)$);
    \draw[dashed] ($(C)+(C|-NULL)$) -- ($(C)+(B|-NULL)$);
    \draw[dashed] ($(C)-(C-|NULL)$) -- ($(C)-(B-|NULL)$);
    \draw[dashed] ($(C)-(C|-NULL)$) -- ($(C)-(B|-NULL)$);
    \draw[dashed] ($3*(C|-NULL)-(NULL|-C)$) -- ($-1*(C)$) -- ($3*(C-|NULL)-(NULL-|C)$) -- ($3*(C)$) -- cycle;

    \draw[fill=red, draw=red] (C) circle (0.1) node[below left] {$(i,j,k)$};
    \draw[fill=red, draw=red] ($(C)+(B-|NULL)$) circle (0.1) node[below right] {$(i,j+1,k)$};
    \draw[fill=red, draw=red] ($(C)+(B|-NULL)$) circle (0.1) node[below right] {$(i+1,j,k)$};
    \draw[fill=red, draw=red] ($(C)-(B|-NULL)$) circle (0.1) node[below left] {$(i-1,j,k)$};
    \draw[fill=red, draw=red] ($(C)-(B-|NULL)$) circle (0.1) node[below right] {$(i,j-1,k)$};
    \draw[fill=red, draw=red] ($-1*(C)$) circle (0.1) node[below left] {$(i-1,j-1,k)$};
    \draw[fill=red, draw=red] ($3*(C)$) circle (0.1) node[below right] {$(i+1,j+1,k)$};
    \draw[fill=red, draw=red] ($3*(C-|NULL)-(NULL-|C)$) circle (0.1) node[below left] {$(i-1,j+1,k)$};
    \draw[fill=red, draw=red] ($3*(C|-NULL)-(NULL|-C)$) circle (0.1) node[below right] {$(i+1,j-1,k)$};

    \draw[red, line width=1pt] ($1.1*(B|-NULL)$) arc (30:60:15);

    \draw[blue, line width=1pt] (B -| D11) -- (D11 -| B) node[at start, below, yshift=-0.2cm] {$\color{blue}\Delta\Gamma_{i,j,k}$};



    
    \coordinate (DIST) at (0.5,0.5);
    \dimline[extension start length=1cm, extension end length=1cm,extension style={black}, label style={above=0.5ex}] {(-0.5,0)}{($(NULL|-B) - (0.5,0)$)}{$A_{i-\frac{1}{2},j,k}=1.0$};
    \dimline[extension start length=-1cm, extension end length=-1cm,extension style={black}, label style={below=0.5ex}] {(0.0,-0.5)}{($(NULL-|B) - (0.0,0.5)$)}{$A_{i,j-\frac{1}{2},k}=1.0$};
    \dimline[extension start length=0.5cm, extension end length=0.5cm,extension style={black}, label style={above=0.5ex}] {($(NULL|-B) + (DIST -| NULL)$)}{($(NULL|-B) + (DIST -| NULL) + (D11|-NULL)$)}{$A_{i,j+\frac{1}{2},k}$};
    \dimline[extension start length=-0.5cm, extension end length=-0.5cm,extension style={black}, label style={below=0.5ex}] {($(NULL-|B) + (DIST |- NULL)$)}{($(NULL-|B) + (DIST|- NULL) + (D11-|NULL)$)}{$A_{i+\frac{1}{2},j,k}$};

    \node[above right] at (NULL) {$\phi > 0 \ \ \alpha_{i,j,k}$};
    \node[below left] at (B) {$\phi < 0$};


    \draw[->, line width=1pt] (-1,-1) -- (-1,0) node[left] {$y$};
    \draw[->, line width=1pt] (-1,-1) -- (0,-1) node[below] {$x$};
    \node[inner sep=2, circle, draw=black, line width=1pt] at (-1,-1) {};
    \node[inner sep=1, circle, draw=none, fill=black] at (-1,-1) {};
    \node[below left] at (-1,-1) {$z$};

\end{tikzpicture}
    \caption{Schematic finite-volume discretization for cut cell $(i,j,k)$ on a Cartesian grid.
    The red dots represent the cell centers. 
    The red line indicates the interface, and the blue line gives the linear approximation of the interface. 
    The fluid with positive level-set values is colored in gray, and the fluid with negative level-set values is colored in white. 
    Volume fraction and apertures are computed for the positive fluid.
    Note that the figure illustrates a two-dimensional slice in the $(x,y)$-plane.
    (For interpretation of the references to color in this figure legend, the reader is referred to the web version of this article.)}
    \label{fig:cut_cell}
\end{figure}
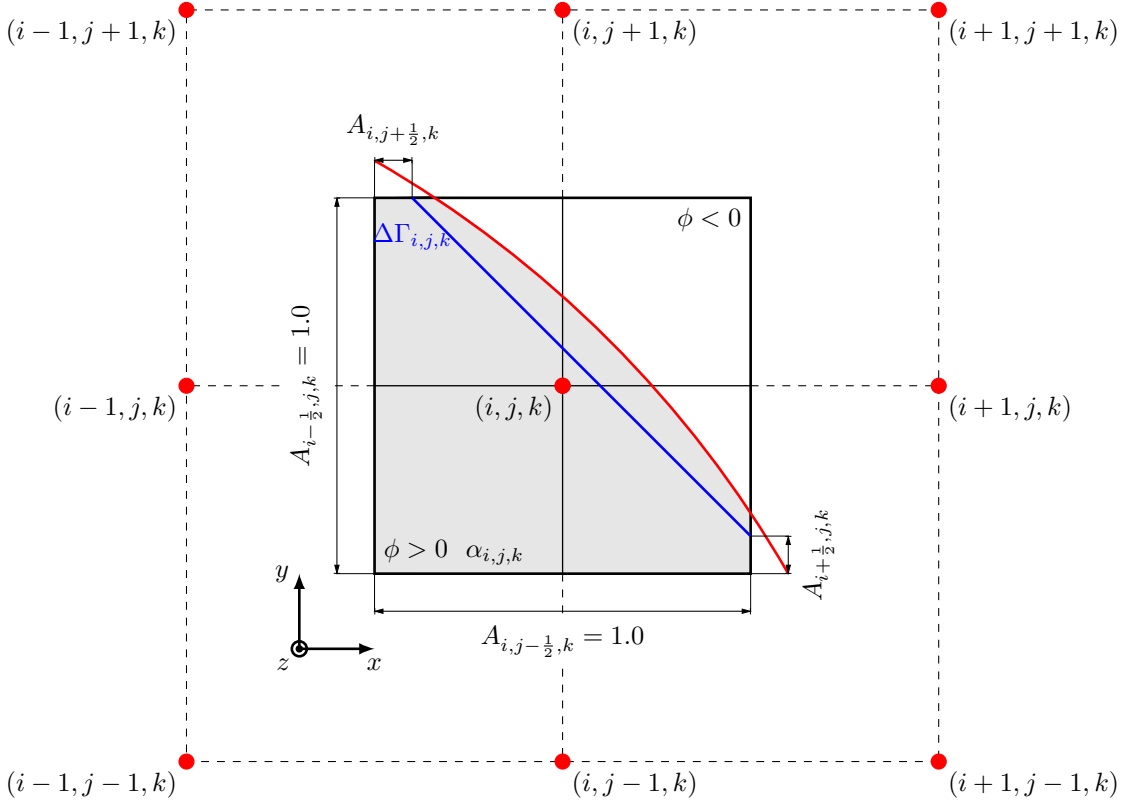

We use the level-set method \cite{Osher1988} to model two-phase flows with fluid-fluid and fluid-solid interfaces.
In particular, we implement the sharp-interface level-set method proposed by Hu et al. \cite{Hu2006}, which
is also used in the solver ALPACA \cite{Hoppe2020a,Hoppe2022a}. 
The interface is tracked by a scalar field $\phi$ whose values represent
the signed distance from the interface of each cell center within the mesh of the finite-volume discretization.
This implies that there is a positive phase ($\phi > 0$) and a negative phase ($\phi < 0$) 
with the interface being located at the zero level-set of $\phi$.
A cell that is intersected by the interface is referred to as cut cell.
Figure \ref{fig:cut_cell} shows a schematic of a cut cell in the finite-volume discretization.
The apertures $A_{i\pm\frac{1}{2},j,k}$, $A_{i,j\pm\frac{1}{2},k}$, and $A_{i,j,k\pm\frac{1}{2}}$ represent the portion of the cell face area that is covered by the respective fluid.
The volume fraction $\alpha_{i,j,k}$ denotes the portion of the cell volume covered by the respective fluid. 
Hereinafter, we will refer to the positive phase with the subscript 1 and to the 
negative phase with the subscript 2. The following relations between the geometrical quantities 
for the positive and negative phase apply: 
\begin{equation}
    \alpha_1 = 1 - \alpha_2,\quad A_1 = 1 - A_2.
    \label{eq:posnegphase}    
\end{equation}
We solve Equation \eqref{eq:FVD} for both phases separately. However, in a cut cell $(i,j,k)$, the equation is modified as follows.
\begin{align}
    \begin{split}
        \frac{\text{d}}{\text{d}t} \alpha_{i,j,k} \bar{\mathbf{U}}_{i,j,k} =
        & - \frac{1}{\Delta x} \left[A_{i+\frac{1}{2},j,k} \left(\mathbf{F}_{i+\frac{1}{2},j,k} + \mathbf{F}^d_{i+\frac{1}{2},j,k}\right) - A_{i-\frac{1}{2},j,k} \left(\mathbf{F}_{i-\frac{1}{2},j,k} + \mathbf{F}^d_{i-\frac{1}{2},j,k} \right) \right] \\
        & - \frac{1}{\Delta y} \left[A_{i,j+\frac{1}{2},k} \left(\mathbf{G}_{i,j+\frac{1}{2},k} + \mathbf{G}^d_{i,j+\frac{1}{2},k}\right) - A_{i,j-\frac{1}{2},k} \left(\mathbf{G}_{i,j-\frac{1}{2},k} + \mathbf{G}^d_{i,j-\frac{1}{2},k} \right) \right] \\
        & - \frac{1}{\Delta z} \left[A_{i,j,k+\frac{1}{2}} \left(\mathbf{H}_{i,j,k+\frac{1}{2}} + \mathbf{H}^d_{i,j,k+\frac{1}{2}}\right) - A_{i,j,k-\frac{1}{2}} \left(\mathbf{H}_{i,j,k-\frac{1}{2}} + \mathbf{H}^d_{i,j,k-\frac{1}{2}} \right) \right] \\
        & + \alpha_{i,j,k} \bar{\mathbf{S}}_{i,j,k}\\
        & - \frac{1}{\Delta x \Delta y \Delta z} \left[\mathbf{X}_{i,j,k}(\Delta \Gamma) + \mathbf{X}^d_{i,j,k}(\Delta \Gamma) \right]
        \label{eq:FVD_levelset}
    \end{split}
\end{align}
The cell-averaged state and the intercell fluxes must be weighted with the volume fraction and the cell face apertures, respectively. The terms $\mathbf{X}(\Delta \Gamma)$
and $\mathbf{X}^d(\Delta \Gamma)$ denote the convective and dissipative interface flux, with $\Delta \Gamma$ being the interface segment length.
We define the projections of the interface segment length on the $x$, $y$, and $z$ direction as the vector 
\begin{equation}
    \Delta \mathbf{\Gamma}_p =
    \begin{pmatrix}
        \Delta \Gamma (\mathbf{i} \cdot \mathbf{n}_I) \\
        \Delta \Gamma (\mathbf{j} \cdot \mathbf{n}_I) \\
        \Delta \Gamma (\mathbf{k} \cdot \mathbf{n}_I) \\
    \end{pmatrix}, \qquad
    \label{eq:interface_segment_projection}
\end{equation}
where $\mathbf{i}$, $\mathbf{j}$, and $\mathbf{k}$ represent the unit vectors in $x$, $y$, and $z$ direction, respectively. The interface normal is given by
$\mathbf{n}_I = \nabla \phi / |\nabla \phi|$.
The interface fluxes read
\begin{equation}
    \mathbf{X} = 
    \begin{pmatrix}
        0 \\
        p_I \Delta \mathbf{\Gamma}_p \\
        p_I \Delta \mathbf{\Gamma}_p \cdot \mathbf{u}_I
    \end{pmatrix}, \qquad
    \mathbf{X}^d = 
    \begin{pmatrix}
        0 \\
        \mathbf{\tau}_I^T \Delta \mathbf{\Gamma}_p \\
        (\mathbf{\tau}_I^T \Delta \mathbf{\Gamma}_p) \cdot \mathbf{u}_I - \mathbf{q}_I \cdot \Delta \mathbf{\Gamma}_p
    \end{pmatrix}.
    \label{eq:interface_flux}
\end{equation}
Here, $p_I$ and $\mathbf{u}_I$ denote the interface pressure and interface velocity. The viscous interface stress tensor $\mathbf{\tau}_I$ is given by
\begin{equation}
    \mathbf{\tau}_I = 
    \begin{pmatrix}
        \tau_I^{11} & \tau_I^{12} & \tau_I^{13} \\
        \tau_I^{21} & \tau_I^{22} & \tau_I^{23} \\
        \tau_I^{31} & \tau_I^{32} & \tau_I^{33} \\
    \end{pmatrix}, \qquad
    \tau_I^{ij} = \mu_I \left(\frac{\partial u_i}{\partial x_j} + \frac{\partial u_j}{\partial x_i}\right) - \frac{2}{3} \mu_I \delta_{ij} \frac{\partial u_k}{\partial x_k}
    , \qquad \mu_I = \frac{\mu_1\mu_2}{\alpha_1\mu_2 + \alpha_2\mu_1}.
\end{equation}
The interface heat flux $\mathbf{q}_I$ reads
\begin{equation}
    \mathbf{q}_I = -\lambda_I \nabla T,
    \qquad \lambda_I = \frac{\lambda_1\lambda_2}{\alpha_1\lambda_2 + \alpha_2\lambda_1}.
\end{equation}
The evaluation of $\mathbf{\tau}_I$ and $\mathbf{q}_I$ requires the computation of velocity and temperature gradients at 
the interface. The gradient at the interface is approximated with the gradient at the cell center.
We use the real fluid state to evaluate these gradients.
The real fluid state in a cut cell is approximated by $\mathbf{W} = \alpha_1\mathbf{W}_1 + \alpha_2\mathbf{W}_2$.
As we solve Equations \eqref{eq:FVD_levelset} and \eqref{eq:interface_segment_projection} for each phase separately,
we emphasize that $\mathbf{n}_I$, $\alpha$, and $A$ must be computed with respect to the present phase. 
Conservation is therefore satisfied since $\mathbf{n}_{I1} = -\mathbf{n}_{I2}$, i.e., the interface flux
terms for the two fluids at the interface have the same magnitude but opposite sign.
The computation of the interface velocity $\mathbf{u}_I$ and interface pressure $p_I$ depends on the type of interface interaction:
\begin{itemize}
    \item For \textbf{fluid-solid} interactions, the interface velocity is prescribed as
    either a constant or a space and time-dependent function. The interface pressure $p_I$ is approximated with the cell pressure.
    \item For \textbf{fluid-fluid} interactions, the two-material Riemann problem at the interface is solved. The solution reads \cite{Hu2004}
    \begin{align}
        \mathbf{u}_I &= \frac{\rho_1c_1 \mathbf{u}_1\cdot \mathbf{n}_{I1} + \rho_2c_2 \mathbf{u}_2\cdot \mathbf{n}_{I1} + p_2 - p_1 - \sigma \kappa}{\rho_1c_1 + \rho_2c_2} \mathbf{n}_{I1}, \notag \\
        p_{I1} &= \frac{\rho_1c_1(p_2 + \sigma \kappa) + \rho_2c_2p_1 + \rho_1c_1\rho_2c_2(\mathbf{u}_2\cdot \mathbf{n}_{I1} - \mathbf{u}_1\cdot \mathbf{n}_{I1})}{\rho_1c_1 + \rho_2c_2}, \\
        p_{I2} &= \frac{\rho_1c_1p_2 + \rho_2c_2(p_1 - \sigma \kappa) + \rho_1c_1\rho_2c_2(\mathbf{u}_2\cdot \mathbf{n}_{I1} - \mathbf{u}_1\cdot \mathbf{n}_{I1})}{\rho_1c_1 + \rho_2c_2}, \notag
    \end{align}
    where $c$ denotes the speed of sound, $\sigma$ is the surface tension coefficient, and $\kappa = \nabla \cdot \mathbf{n}_{I1}$ denotes the curvature. 
    For $\sigma \neq 0$ and $\kappa \neq 0$, the interface pressure experiences a jump at the interface as constituted by mechanical equilibrium. The interface pressure 
    in Equation \eqref{eq:interface_flux} must be chosen with respect to the present phase.
\end{itemize}

\begin{figure}[!h]
    \centering
    \begin{tikzpicture}
        \draw[thick, black] (0,0) -- (0,2) node[at end, circle, draw=blue] {} -- (2,2) node[at end, circle, draw=blue] {} -- (2,0) node[at end, circle, draw=blue] {} -- (0,0) node[at end, circle, draw=blue, fill=blue] {};
        \draw[thick, dashed] (1.0,0.0) -- (0.0,1.0);
        \draw[thick, black] (3,0) -- (3,2) node[at end, circle, draw=blue, fill=blue] {} -- (5,2) node[at end, circle, draw=blue] {} -- (5,0) node[at end, circle, draw=blue] {} -- (3,0) node[at end, circle, draw=blue, fill=blue] {};
        \draw[thick, dashed] (3.8,0.0) -- (4.2,2.0);
        \draw[thick, black] (6,0) -- (6,2) node[at end, circle, draw=blue, fill=blue] {} -- (8,2) node[at end, circle, draw=blue] {} -- (8,0) node[at end, circle, draw=blue, fill=blue] {} -- (6,0) node[at end, circle, draw=blue] {};
        \draw[thick, dashed] (6.0,0.5) -- (7.0,2.0);
        \draw[thick, dashed] (7.2,0.0) -- (8.0,0.8);
    \end{tikzpicture}
    \caption{Three typical cut cell face configurations.
    Solid and hollow blue circles indicate positive and negative level-set corner values, respectively.}
    \label{fig:cut_cell_face}
\end{figure}
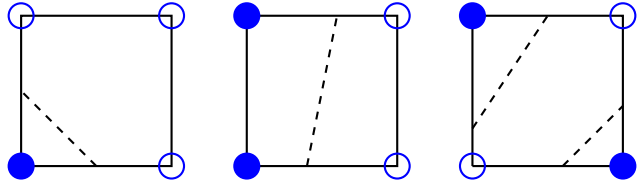

Assuming a linear interface within each cut cell, the cell face apertures are computed analytically as follows.
The level-set values at the edges of a computational cell are computed using trilinear 
interpolation. The sign of the level-set values at the four corners of a cell face
determine the cut cell face configuration. Figure \ref{fig:cut_cell_face} illustrates three typical cases of a cell face that is intersected by the interface.
In total, there are $2^4$ different sign combinations of the level-set values along the corners. Hence, there are $2^4$ different cut cell face configurations. 
For each of these, the cell face aperture is evaluated as the (sum of) areas of the basic geometric shapes, i.e., triangle or trapezoid.
The interface segment length $\Delta \Gamma$ is computed from the apertures as follows.
\begin{equation}
    \Delta \Gamma_{i,j,k} = \left[(A_{i+\frac{1}{2},j,k} - A_{i-\frac{1}{2},j,k})^2\Delta y\Delta z + (A_{i,j+\frac{1}{2},k} - A_{i,j-\frac{1}{2},k})^2\Delta x\Delta z  + (A_{i,j,k+\frac{1}{2}} - A_{i,j,k-\frac{1}{2}})^2\Delta x\Delta y  \right]^\frac{1}{2}
\end{equation}
Geometrical reconstruction with seven pyramids yields the volume fraction $\alpha$.
\begin{align}
    \alpha_{i,j,k} &= \frac{1}{3} \frac{1}{\Delta x \Delta y \Delta z} \left[ A_{i+\frac{1}{2},j,k} \Delta y \Delta z \frac{1}{2} \Delta x +  A_{i-\frac{1}{2},j,k} \Delta y \Delta z \frac{1}{2} \Delta x  + A_{i,j+\frac{1}{2},k} \Delta x \Delta z \frac{1}{2} \Delta y \right. \notag \\
    &+ \left. A_{i,j-\frac{1}{2},k} \Delta x \Delta z \frac{1}{2} \Delta y + A_{i,j,k-\frac{1}{2}} \Delta x \Delta y \frac{1}{2} \Delta z + A_{i,j,k+\frac{1}{2}} \Delta x \Delta y \frac{1}{2} \Delta z + \Delta \Gamma_{i,j,k} \phi_{i,j,k}  \vphantom{\frac12}\right]
\end{align}
Note that the described approach yields the volume fraction and apertures with respect to the positive phase.
The values of the negative phase can be obtained from relations \eqref{eq:posnegphase}.

The level-set field is advected by the interface velocity $\mathbf{u}_I= \mathbf{n}_I u_I$ by solving the level-set advection equation. 
\begin{equation}
    \frac{\partial \phi}{\partial t} + \mathbf{u}_I\cdot\nabla\phi = 0
    \label{eq:levelset_advection}
\end{equation}
The spatial term in Equation \eqref{eq:levelset_advection} is discretized using high-order upstream central (HOUC) \cite{Nourgaliev2007} stencils.
For the temporal integration, we apply the same scheme that is used to integrate the conservative variables, which typically is a Runge-Kutta method, see Subsection \ref{subsec:timeint}.
The level-set field is only advected within a narrowband around the interface.

In order to apply the reconstruction stencils used in the finite-volume discretization near the interface,
we extrapolate the real fluid state to the other side of the interface.
We denote the cells on the other side of the interface as ghost cells. 
An arbitrary quantity $\psi$ is extrapolated from the real cells to the ghost cells
by advancing it in a fictitious time $\tau$ to steady state according to
\begin{equation}
    \frac{\partial \psi}{\partial \tau} \pm \mathbf{n}_I\cdot\nabla \psi = 0.
    \label{eq:extension}
\end{equation}
The sign of the interface normal $\pm \mathbf{n}_I$
depends on the sign of the present phase: To extrapolate the real fluid state of the positive phase into its ghost cells, we must extrapolate 
into the direction of the negative level-set, i.e. $-\mathbf{n}_I$, and vice versa.
The spatial term is discretized using a first-order upwind stencil. 
Temporal integration is performed with the Euler method.
Note that we also use this extension procedure to extend the interface velocity from the cut cells into the narrowband around the interface,
where we advect the level-set field.

The computation of the geometrical quantities requires the level-set field to be a signed distance function. During a simulation,
the level-set field loses its signed distance property due to numerical errors and/or a shearing flow field. Additionally, since we only advect
the level-set field within a narrowband around the interface, the level-set field develops a kink at the edge of the narrowband and the 
remainder of the computational domain.
The signed distance property is maintained via reinitialization of the level-set field. The reinitialization equation reads
\begin{equation}
    \frac{\partial \phi}{\partial \tau} + sgn(\phi^0) (|\nabla\phi| - 1) = 0.
    \label{eq:reinitialization}
\end{equation}
Here, $\phi^0$ represents the level-set at a fictitious time $\tau=0$. We apply first-order \cite{Russo2000} or higher-order WENO-HJ \cite{jiang2000weighted} schemes
to solve this equation. We reinitialize the level-set each physical time step for a fixed amount of fictitious time steps resulting
in a sufficiently small residual of Equation \eqref{eq:reinitialization}.

The presented level-set method is not consistent when the interface crosses a cell face within a single time step,
i.e., when new fluid cells are created or fluid cells have vanished. 
Figure \ref{fig:mixing} displays this scenario for a 1D discretization. The interface 
is moving from cell $i-1$ to cell $i$. At $t^n$, cell $i$ is a newly created cell w.r.t. phase 1 and cell $i-1$ is a vanished cell w.r.t. phase 2.
To maintain conservation, we must do the following.
For phase 1, conservative variables must be taken from cell $i-1$ and put into the newly created cell $i$.
For phase 2, conservative variables must be taken from cell $i$ and put into the vanished cell $i-1$.
In addition to the scenario where an interface crosses the cell face, small cut cells may generally lead to an unstable
integration using the time step restriction that is based on a full cell. 
We apply a mixing procedure \cite{Hu2006} that deals with these problems. The procedure is applied to two types of cells. 
\begin{enumerate}
    \item Cells where $\alpha = 0$ after integration but $\alpha \neq 0$ before (vanished cells). 
    \item Cells with $\alpha < \alpha_{\text{mix}}$ after integration (newly created cells and small cells). 
\end{enumerate}
We use a mixing threshold of $\alpha_{\text{mix}} = 0.6$. For each cell that requires mixing, we identify target (trg) cells from the interface normal.
Consider cell $i$ in Figure \ref{fig:mixing}, which is a small/newly created
cell for phase 1. Here, the target cell in $x$ direction is cell $i-1$, as $\mathbf{n}_{I1}\cdot \mathbf{i} < 0$. Analogously, 
cell $i-1$ is a vanished cell for phase 2. The corresponding target is cell $i$, since $\mathbf{n}_{I2}\cdot \mathbf{i} > 0$. 
In 3D, there are 7 target cells in total:
One in each spatial direction $x$, $y$, and $z$, one in each plane $xy$, $xz$, and $yz$, and one in $xyz$.
Seven mixing weights are computed as
\begin{align}
    \beta_x &= |\mathbf{n}_I\cdot \mathbf{i}|^2 \alpha_{trg}, \notag \\
    \beta_y &= |\mathbf{n}_I\cdot \mathbf{j}|^2 \alpha_{trg}, \notag \\
    \beta_z &= |\mathbf{n}_I\cdot \mathbf{k}|^2 \alpha_{trg}, \notag \\
    \beta_{xy} &= |\left(\mathbf{n}_I\cdot \mathbf{i}\right) \left(\mathbf{n}_I\cdot \mathbf{j}\right) | \alpha_{trg}, \\
    \beta_{xz} &= |\left(\mathbf{n}_I\cdot \mathbf{i}\right) \left(\mathbf{n}_I\cdot \mathbf{k}\right) | \alpha_{trg}, \notag \\
    \beta_{yz} &= |\left(\mathbf{n}_I\cdot \mathbf{j}\right) \left(\mathbf{n}_I\cdot \mathbf{k}\right) | \alpha_{trg}, \notag \\
    \beta_{xyz} &= |\left(\mathbf{n}_I\cdot \mathbf{i}\right) \left(\mathbf{n}_I\cdot \mathbf{j}\right) \left(\mathbf{n}_I\cdot \mathbf{k}\right) |^{2/3} \alpha_{trg}. \notag
\end{align}
Here, $\alpha_{trg}$ denotes the volume fraction of the target cell in the corresponding direction.
We normalize the mixing weights so that $\sum_{trg} \beta_{trg} =1$, where $trg\in\{x,y,z,xy,xz,yz,xyz\}$.
Subsequently, the mixing flux $\mathbf{M}_{trg}$ is computed like
\begin{equation}
    \mathbf{M}_{trg} = \frac{\beta_{trg}}{\alpha \beta_{trg} + \alpha_{trg}} \left[(\alpha_{trg}\bar{\mathbf{U}}_{trg})\alpha - (\alpha \bar{\mathbf{U}})\alpha_{trg} \right].
\end{equation}
The conservative variables are then updated according to
\begin{align}
    \alpha \bar{\mathbf{U}} &= \left( \alpha \bar{\mathbf{U}} \right)^* + \sum_{trg}\mathbf{M}_{trg}, \\
    \alpha_{trg} \bar{\mathbf{U}}_{trg} &= \left( \alpha_{trg} \bar{\mathbf{U}}_{trg} \right)^* - \sum_{trg}\mathbf{M}_{trg}.
\end{align}
Here, $\alpha \bar{\mathbf{U}}$ and $\alpha_{trg} \bar{\mathbf{U}}_{trg}$ denote the conservative variables of the cells that require mixing 
and the conservative variables of the corresponding target cells, respectively.
Star-quantities denote conservative variables before mixing.
\begin{figure}
    \centering
    \begin{tikzpicture}
        \draw[line width=1pt] (0,0) -- (14,0);
        \node[circle, draw=black, fill=black, inner sep=1pt] (A) at (4,0) {};
        \node[circle, draw=black, fill=black, inner sep=1pt] (B) at (10,0) {};
        \node[below,yshift=-0.5cm] at (A) {$i-1$};
        \node[below,yshift=-0.5cm] at (7,0) {$i-\frac{1}{2}$};
        \node[below,yshift=-0.5cm] at (B) {$i$};
        \node[color=blue] at ([yshift=2.2cm]A) {vanished cell};
        \node[color=blue] at ([yshift=2.2cm]B) {target cell};
        \node[color=red] at ([yshift=-2.2cm]A) {target cell};
        \node[color=red] at ([yshift=-2.2cm]B) {newly created cell};
        \node[color=red] at (1,-2.2) {$\phi > 0$};
        \node[color=blue] at (13,2.2) {$\phi < 0$};

        \draw (1,0.3) -- (1,-0.3);
        \draw (7,0.3) -- (7,-0.3);
        \draw (13,0.3) -- (13,-0.3);
        \draw[line width=0.8pt, dashed, color=green!50!black] (6.2,1.5) -- (6.2,-1.5) node[at start, above] {$\Gamma_0(t^{n-1})$};
        \draw[line width=0.8pt, dashed, color=green!50!black] (8.5,1.5) -- (8.5,-1.5) node[at start, above] {$\Gamma_0(t^n)$};
        \draw[<->] (7,-1.5) -- (8.5,-1.5) node[above, midway] {$\alpha_1$};
        \draw[<->] (8.5,1.5) -- (13,1.5) node[below, midway] {$\alpha_2$};

        \draw[->, line width=1pt] (5,-3) -- (9,-3) node[below, yshift=-0.2cm, midway] {Mixing flux $\mathbf{M}_1$};
        \draw[<-, line width=1pt] (5,3) -- (9,3) node[above, yshift=0.2cm, midway] {Mixing flux $\mathbf{M}_2$};
    \end{tikzpicture}
    \caption{Schematic illustrating the mixing procedure in a 1D discretization at $t^n$. Red and blue color indicate the positive and negative phases.
    Green indicates interface positions $\Gamma_0$ at $t^{n-1}$ and $t^n$.}
    \label{fig:mixing}
\end{figure}

\subsection{Computational Domain and Boundary Conditions}

The computational domain is a cuboid. Figure \ref{fig:computational_domain} depicts an
exemplary computational domain including the nomenclature for the boundary locations.
The solver provides symmetry, periodic, no-slip wall, Dirichlet, and Neumann boundary conditions. 
The no-slip wall boundary condition allows the user to specify either a constant value or time-dependent function for the wall velocity. 
The Dirichlet and Neumann boundary conditions allow the user to specify either a constant value or a space and time-dependent function. 
Furthermore, (in 2D only) multiple different types of boundary conditions may be imposed along a single boundary location.
Here, the user must specify the starting and end point of each of the different boundary types along the specific boundary location.
The level-set implementation allows for arbitrary immersed solid boundaries.

\begin{figure}
    \centering
    \begin{tikzpicture}
        \node (-0.1, -0.1) {};
        \draw[fill=blue, opacity=0.1] (0,0) -- (0,3) -- (5,3) -- (5,0) -- cycle;
        \draw[fill=blue, opacity=0.1] (1.5,1.5) -- (1.5,4.5) -- (6.5,4.5) -- (6.5,1.5) -- cycle;
        \draw[fill=red, opacity=0.1] (5,0) -- (5,3) -- (6.5,4.5) -- (6.5,1.5) -- cycle;
        \draw[fill=red, opacity=0.1] (0,0) -- (0.0,3.0) -- (1.5,4.5) -- (1.5,1.5) -- cycle;
        \draw[fill=green, opacity=0.1] (0,0) -- (1.5,1.5) -- (6.5,1.5) -- (5,0) -- cycle;
        \draw[fill=green, opacity=0.1] (0,3) -- (1.5,4.5) -- (6.5,4.5) -- (5,3) -- cycle;
        \draw[thick] (0,0) -- (0,3) -- (5,3) -- (5,0) -- cycle;
        \draw[thick] (0,3) -- (1.5,4.5) -- (6.5,4.5) -- (5,3);
        \draw[thick] (5,0) -- (6.5,1.5) -- (6.5,4.5);
        \node[opacity=0.5] at (3.25,0.75) {\textcolor{green!70!black}{south}};

        \node[opacity=0.5] at (4,2.8) {\textcolor{blue!70!black}{top}};
        \draw[draw=none] (0.0,1.5) -- (1.5,3.0) node[midway, sloped] {\textcolor{red!70!black}{west}};
        \draw[draw=none] (5.0,1.5) -- (6.5,3.0) node[midway, sloped] {\textcolor{red!70!black}{east}};
        \draw[line width=1pt, ->] (1.5,1.5) -- (2.2,1.5) node[at end, above] {$x$};
        \draw[line width=1pt, ->] (1.5,1.5) -- (1.5,2.2) node[at end, left] {$y$};
        \draw[line width=1pt, ->] (1.5,1.5) -- (1.1,1.1) node[at end, above left] {$z$};
        \node at (3.25,3.75) {\textcolor{green!70!black}{north}};
        \node at (2.5,1.3) {\textcolor{blue!70!black}{bottom}};
    \end{tikzpicture}
    \caption{Computational domain with boundary locations.}
    \label{fig:computational_domain}
\end{figure}

\begin{table}[t!]
    \begin{center}
        \footnotesize
        \begin{tabular}{c c c} 
        \hline
        Time Integration    & Euler & \\
                            & TVD-RK2 \cite{Gottlieb1998a} & \\
                            & TVD-RK3 \cite{Gottlieb1998a} & \\
        \hline

        Flux Function/Riemann Solver    & Lax-Friedrichs (LxF) & According to \\
                                        & Local Lax-Friedrichs (LLxF, Rusanov) & According to \\
                                        & HLL/HLLC/HLLC-LM \cite{Harten1983a,Toro1994,Toro2009a,Toro2019,Fleischmann2020} & Signal speed estimates see below \\
                                        & AUSM+ \cite{Liou1996} & \\     
                                        & Componentwise LLxF & Flux-splitting formulation \\
                                        & Roe \cite{Roe1981} & Flux-splitting formulation \\
        \hline
        
        Signal Speed Estimates  & Arithmetic    & \\
                                & Davis \cite{Davis1988}         & \\
                                & Einfeldt \cite{Einfeldt1988a}     & \\
                                & Toro \cite{Toro1994}  & \\

        \hline

        Spatial Reconstruction  & WENO1 \cite{Jiang1996} & \\
                                & WENO3-JS/Z/N/F3+/NN \cite{Jiang1996,Acker2016a,Gande2020,Bezgin2021b} & \\
                                & WENO5-JS/Z \cite{Jiang1996,Borges2008a} & \\
                                & WENO6-CU/CUM  \cite{Hu2010,Hu2011} & \\
                                & WENO7-JS \cite{Balsara2000}& \\
                                & WENO9-JS \cite{Balsara2000}& \\
                                & TENO5 \cite{Fu2016} & \\
                                & Second-order central & For dissipative terms only \\
                                & Fourth-order central & For dissipative terms only \\

        \hline

        Spatial Derivatives & Second-order central & \\
                            & Fourth-order central & \\
                            & HOUC-3/5/7 \cite{Nourgaliev2007}& \\
        \hline
        Levelset reinitialization  & First-order \cite{Russo2000} & \\
                                   & HJ-WENO \cite{jiang2000weighted} & \\

        \hline
        Ghost fluid extension & First-order upwind \cite{Hu2006} & \\

        \hline

        LES Modules & ALDM \cite{Hickel2014b} & \\

        \hline 

        Equation of State   & Ideal Gas & \\
                            & Stiffened Gas \cite{Menikoff1989} & \\
                            & Tait \cite{Fedkiw1999a}& \\

        \hline Boundary Conditions  & Periodic & \\
                                    & Zero Gradient & E.g., used for outflow boundaries \\
                                    & Neumann & E.g., for a prescribed heat flux \\
                                    & Dirichlet & \\
                                    & No-slip Wall &  \\
                                    & Immersed Solid Boundaries & Arbitrary geometries via level-set \\

        \hline
    \end{tabular}
    \caption{Overview on numerical methods available in JAX-FLUIDS.}
    \label{tab:NumericalMethods}
    \end{center}
\end{table}
\section{Software Implementation Details}
\label{sec:Implementation}

In the past, CFD solvers have been written predominantly in low-level programming languages like Fortran and C/C++. These languages
offer computational performance and CPU parallelization capabilities. However, the integration of ML models, which are typically coded in Python,
is not straightforward and automatic differentiation capabilities are nonexistent. With the present work, we want to provide a CFD framework that
allows for seamless integration and end-to-end optimization of ML models, without major performance losses.
The Python package JAX \cite{jax2018github} satisfies these requirements.
Therefore, we use JAX as the fundamental building block of JAX-FLUIDS.
In this section, we give implementation details and the algorithmic structure of JAX-FLUIDS.

\subsection{Array Programming in JAX}

JAX is a Python library for high-performance numerical computations, which uses XLA to compile and run code on accelerators like CPUs, GPUs, and TPUs.
Designed as a machine learning library, JAX supports automatic differentiation \cite{Baydin2018}.
In particular, JAX comes with a fully-differentiable version of the popular NumPy package \cite{Harris2020} called JAX NumPy.
Both Python and NumPy are widely popular and easy to use. NumPy adds high-level functionality to handle large arrays and matrices.

The backbone of (JAX) NumPy is the multidimensional array, so called \mintinline{python}{jax.numpy.DeviceArray}.
We use arrays to store all of our field data.
In particular, the vectors of conservative and primitive variables, $\mathbf{U}$ and $\mathbf{W}$, are stored in arrays of shape $(5, N_x + 2 N_h, N_y + 2 N_h, N_z + 2 N_h)$.
$(N_x, N_y, N_z)$ denote the resolution in the three spatial directions, and $N_h$ specifies the number of halo cells.
Our implementation naturally degenerates for one- and two-dimensional settings by using only a single cell in the excess dimensions.

In the \textit{array programming paradigm} (see \cite{Walt2011} for details on numerical operations on arrays in NumPy), every operation is performed on the entire array.
I.e., instead of writing loops over the array as is common in Fortran or C/C++, we use appropriate indexing/slicing operations.
As a result, many parts of the code used from classical CFD solvers have to be rewritten according to the \textit{array programming paradigm}.
As an example we include the source code of our implementation of the second-order central cell face reconstruction scheme in Figure \ref{fig:reconstruction}.
The \mintinline{python}{class CentralSecondOrderReconstruction} inherits from \mintinline{python}{class SpatialReconstruction}. This parent class
has an abstract member function \mintinline[escapeinside=||,mathescape=true]{python}{reconstruct_xi}, which must be implemented in every child class. 
\mintinline[escapeinside=||,mathescape=true]{python}{reconstruct_xi} receives the entire \mintinline{python}{buffer} and the 
reconstruction direction \mintinline{python}{axis} as arguments.

The data buffer has to be indexed/sliced differently depending on the spatial direction of the reconstruction and the dimensionality of the problem.
Note that the buffer array has shape $(5, N_x + 2 N_h, N_y + 2 N_h, N_z + 2 N_h)$ if the problem is three-dimensional, 
$(5, N_x + 2 N_h, N_y + 2 N_h, 1)$ if the problem is two-dimensional, and $(5, N_x + 2 N_h, 1, 1)$ if the problem is one-dimensional. 
The slice indices have to be adapted accordingly.
To prevent boilerplate code in the reconstruction routine, we abstract the slicing operations.
The member variable \mintinline{python}{self.slices} is a list that holds the correct slice objects for each spatial direction.
Consider reconstructing in $x$-direction (\mintinline{python}{axis=0}) using the present second-order cell face 
reconstruction scheme,
\begin{equation}
    \mathbf{U}_{i+\frac{1}{2},j,k}= \frac{1}{2}( \mathbf{U}_{i,j,k} + \mathbf{U}_{i+1,j,k} ).
\end{equation}
Here, we require two slice objects: \mintinline{python}{jnp.s_[...,self.nh-1:-self.nh,self.nhy,self.nhz]} for $\mathbf{U}_{i,j,k}$ and
\mintinline{python}{jnp.s_[...,self.nh:-self.nh+1,self.nhy,self.nhz]} for $\mathbf{U}_{i+1,j,k}$.
The variable \mintinline{python}{self.nh} denotes the number of halo cells.
The slices in $x$-direction, i.e., \linebreak \mintinline{python}{self.nh-1:-self.nh} and \mintinline{python}{self.nh:-self.nh+1}, are determined by the
reconstruction scheme itself. 
\mintinline{python}{self.nhy} and \mintinline{python}{self.nhz} denote the slice objects for the dimensions
in which we do not reconstruct. These are either \mintinline{python}{self.nh:-self.nh} if the dimension is active
or \mintinline{python}{None:None} if the dimension is inactive. \mintinline{python}{self.nhx}, \mintinline{python}{self.nhy}, and \mintinline{python}{self.nhz}
are defined in the parent class.

\begin{figure}
    \centering
\begin{minted}{python}
from functools import partial
import jax, jax.numpy as jnp
from jaxfluids.stencils.spatial_reconstruction import SpatialReconstruction

class CentralSecondOrderReconstruction(SpatialReconstruction):

    def __init__(self, nh: int, inactive_axis: jnp.array) -> None:
        super(CentralSecondOrderReconstruction, self).__init__(nh=nh, inactive_axis=inactive_axis)
        self.slices = [
            [   jnp.s_[..., self.nh-1:-self.nh  , self.nhy, self.nhz],         #  X-DIRECTION          
                jnp.s_[..., self.nh  :-self.nh+1, self.nhy, self.nhz],   ],

            [   jnp.s_[..., self.nhx, self.nh-1:-self.nh  , self.nhz],         #  Y-DIRECTION   
                jnp.s_[..., self.nhx, self.nh  :-self.nh+1, self.nhz],   ],   

            [   jnp.s_[..., self.nhx, self.nhy, self.nh-1:-self.nh  ],         #  Z-DIRECTION 
                jnp.s_[..., self.nhx, self.nhy, self.nh  :-self.nh+1],   ],   
        ]

    @partial(jax.jit, static_argnums=(0, 2))
    def reconstruct_xi(self, buffer: jnp.array, axis: int) -> jnp.array:
        s_ = self.slices[axis]
        cell_face_state_xi = 0.5 * ( buffer[s_[0]] + buffer[s_[1]] )
        return cell_face_state_xi
\end{minted}
    \caption{Code snippet for the second-order central cell face reconstruction stencil.}
    \label{fig:reconstruction}
\end{figure}

\subsection{Object-oriented Programming in the Functional Programming World of JAX}

As already alluded to in the previous subsection, we use object-oriented programming (OOP) throughout the entire JAX-FLUIDS solver.
Although JAX leans inherently more towards a functional programming style, we have opted to program JAX-FLUIDS in a modular object-oriented approach
since this has several benefits.
Firstly, comprehensive CFD solvers typically offer a plethora of interchangeable numerical algorithms.
Naturally, OOP allows the implementation of many derived classes, e.g., different spatial reconstructors
(see \mintinline{python}{class SpatialReconstruction} in the previous subsection), and saves boilerplate code. 
Secondly, the modular OOP approach allows users to customize a solver specific to their problem.
Via a numerical setup file, the user can detail the numerical setup prior to every simulation.
Thirdly, we want to stress that the modularity of our solver allows for straightforward integration of custom modules and implementations.
For example, avid ML-CFD researchers can easily implement their own submodule into the JAX-FLUIDS framework, either simply for forward simulations or for learning new routines from data.

\subsection{Just-in-time (jit) Compilation and Pure Functions}
JAX offers the possibility to just-in-time (jit) compile functions, which significantly increases the performance. However,
jit-compilation imposes two constraints:
\begin{enumerate}
    \item \textbf{The function must be a pure function}. A function is pure if its return values are identical for identical input arguments and the function has no side effects. 
    \item \textbf{Control flow statements in the function must not depend on input argument values}. During jit-compilation, an abstract version of the function is cached that works for arbitrary argument values.
    I.e., the function is not compiled for concrete argument values but rather for the set of all possible argument values
    where only the array shape and type is fixed. Therefore, control flow statements that depend on the input argument values can not be jit-compiled, unless 
    the argument is a \textit{static argument}. In this case, the function is recompiled for all values that the \textit{static argument} takes during runtime.
\end{enumerate}
The aforementioned requirements to jit-compile a function have the following impact on our implementation:
\begin{itemize}
    \item The \mintinline{python}{self} argument in jit-compiled member functions must be a \textit{static argument}. 
    This implies that class member variables that are used in the function are static and hence must not be modified. 
    In other words, the type of class member variables is similar to the C++ type \mintinline{C++}{constexpr}.
    \item Control flow statements in jit-compiled functions can only be evaluated on \textit{static arguments}. 
    In our code, there are three distinct types of control flow statements where this has an impact:
    \begin{enumerate}
        \item \textbf{Conditional exit of a for/while loop}.
        The top level compute loop over the physical simulation time (compare Algorithm \ref{alg:main_loop}) is not jit-compiled, since it consists of a while loop that is 
        exited when the final simulation time is reached $t \geq t_{end}$. However, $t$ is not a static variable. 
        We therefore only jit-compile the functions \mintinline{python}{compute_timestep}, \\ \mintinline{python}{do_integration_step()}, and \mintinline{python}{compute_forcings()}. 
        These are the functions that constitute the heavy compute within the main loop. 
        \item \textbf{Conditional slicing of an array}.
        In multiple parts of the JAX-FLUIDS code, arrays are sliced depending on the present spatial direction.
        The present spatial direction is indicated by the input argument \mintinline{python}{axis} (compare cell face reconstruction in Figure \ref{fig:reconstruction}).
        The code that is actually executed conditionally depends on the value of \mintinline{python}{axis}. Therefore, \mintinline{python}{axis} must be a static argument. 
        Functions that receive \mintinline{python}{axis} as an input argument are compiled for all values that \mintinline{python}{axis} might take during runtime.
        Each compiled version of those functions is cached.
        \item \textbf{Conditional executation of code sections}. 
        We explained above that class member variables are always static. 
        In practice, we often use them to conditionally compile code sections, 
        much like the \mintinline{C++}{if constexpr} in C++. 
        An example is the cell face reconstruction of the variables, which can be done on primitive or conservative variables 
        in either physical or characteristic space. 
    \end{enumerate}
    \item Element-wise conditional array operations are implemented using masks, as jit-compilation requires the array shapes to be fixed at compile time.
    Figure \ref{fig:masks} exemplarily illustrates the use of masks, here for the evaluation of the level-set advection equation.
    We make frequent use of \mintinline{python}{jnp.where} to implement element-wise conditional array operations.
\end{itemize}

\begin{figure}
    \centering
\begin{minted}{python}
from functools import partial
import jax, jax.numpy as jnp

class LevelsetHandler():
    def __init__(self, ...) -> None:
        pass

    @partial(jax.jit, static_argnums=(0, 3))
    def compute_levelset_advection_rhs_xi(self, levelset: jnp.array, interface_velocity: jnp.array,
            axis: int) -> jnp.array:

        # LEFT AND RIGHT SIDED DERIVATIVE
        derivative_L = self.spatial_derivative.derivative_xi(levelset, self.cell_sizes[axis], axis, 0)
        derivative_R = self.spatial_derivative.derivative_xi(levelset, self.cell_sizes[axis], axis, 1)

        # UPWINDING DEPENDING ON LOCAL INTERFACE VELOCITY
        velocity    = interface_velocity[axis]
        mask_L      = jnp.where(velocity >= 0.0, 1.0, 0.0)
        mask_R      = 1.0 - mask_L

        # RIGHT-HAND-SIDE EVALUATION
        rhs_contribution = - velocity * (mask_L * derivative_L + mask_R * derivative_R)

        return rhs_contribution
\end{minted}
    \caption{Code snippet showing the right-hand-side computation of the level-set advection equation. }
    \label{fig:masks}
\end{figure}

\subsection{Main Objects and Compute Loops}
\label{subsec:ComputeLoops}

We put emphasis on making JAX-FLUIDS an easy-to-use Python package for ML-CFD research. 
Figure \ref{fig:run_solver} shows the required code lines to run a simulation.
The user must provide a numerical setup and a case setup file in the \textit{json} format.
The numerical setup specifies the combination of numerical methods that will be used for the simulation.
The case setup details the physical properties of the simulation including the spatial domain and its resolution, initial and boundary conditions, and material properties.
As an example, we include the numerical setup and case setup for the Sod shock tube test case in the appendix
in Figure \ref{fig:numerical_setup} and \ref{fig:case_setup}, respectively.
Using the \textit{json} files, we create an \mintinline{python}{class InputReader} instance, which we denote as \mintinline{python}{input_reader}.
The \mintinline{python}{input_reader} performs necessary data type transformations and a thorough sanity check ensuring 
that the provided numerical and case setups are consistent. 
Then, an \mintinline{python}{class Initializer} and a \mintinline{python}{class SimulationManager} object are created 
using the \mintinline{python}{input_reader} object.
The \mintinline{python}{class Initializer} implements functionality to generate the initial buffers
from either the initial condition specified in the case setup file or from a restart file. We receive these buffers 
by calling the \mintinline{python}{initialization} method of the \mintinline{python}{initializer} object.
The \mintinline{python}{class SimulationManager} is the main class in JAX-FLUIDS, implementing the algorithm to advance the
initial buffers in time using the specified numerical setup. The initial buffers
must be passed to the \mintinline{python}{simulate} method of the \mintinline{python}{simulation_manager} object.
The simulation starts upon execution of this function.

\begin{figure}
\begin{minted}{python}
import json
from jaxfluids import InputReader, Initializer, SimulationManager

numerical_setup_dict    = json.load(open("numerical_setup.json"))
case_setup_dict         = json.load(open("case_setup.json"))

input_reader        = InputReader(case_setup_dict, numerical_setup_dict)
initializer         = Initializer(input_reader)
simulation_manager  = SimulationManager(input_reader)

initial_buffer      = initializer.initialization()
simulation_manager.simulate(initial_buffer)
\end{minted}
    \caption{Code snippet illustrating how to run a simulation with JAX-FLUIDS.}
    \label{fig:run_solver}
\end{figure}
    
The code contains three major loops:
\begin{enumerate}
    \item \textbf{Loop over physical simulation time}, see Algorithm \ref{alg:main_loop}.
    \item \textbf{Loop over Runge-Kutta stages}, see Algorithm \ref{alg:do_integration_step}.
    \item \textbf{Loop over active spatial dimensions}, see Algorithm \ref{alg:compute_rhs}.
\end{enumerate}

\begin{algorithm}
\caption{Loop over physical simulation time in the \mintinline[escapeinside=||,mathescape=true]{python}{simulate} function.
Red text color indicates functions that are only executed for simulations with active forcings.}
\label{alg:main_loop}
\While{time < end time}{
    \mintinline[escapeinside=||,mathescape=true]{python}{compute_timestep()}\\
    \textcolor{red}{\mintinline[escapeinside=||,mathescape=true]{python}{forcings_handler.compute_forcings()}}\\
    \mintinline[escapeinside=||,mathescape=true]{python}{do_integration_step()}\\
    \mintinline[escapeinside=||,mathescape=true]{python}{output_writer.write_output()}
}
\end{algorithm}

\begin{algorithm}
    \caption{Loop over Runge-Kutta stages in the \mintinline[escapeinside=||,mathescape=true]{python}{do_integration_step} function.
    Blue text color indicates functions that are only executed for two-phase simulations.}
    \label{alg:do_integration_step}
    \For{RK stages}{
        \mintinline[escapeinside=||,mathescape=true]{python}{space_solver.compute_rhs()}\\
        \textcolor{blue}{\mintinline[escapeinside=||,mathescape=true]{python}{levelset_handler.transform_volume_averages_to_conservatives()}\\}
        \mintinline[escapeinside=||,mathescape=true]{python}{time_integrator.prepare_buffers_for_integration()}\\
        \mintinline[escapeinside=||,mathescape=true]{python}{time_integrator.integrate_conservatives()}\\
        \textcolor{blue}{\mintinline[escapeinside=||,mathescape=true]{python}{time_integrator.integrate_levelset()}}\\
        \textcolor{blue}{\mintinline[escapeinside=||,mathescape=true]{python}{levelset_handler.reinitialize_levelset()}}\\
        \textcolor{blue}{\mintinline[escapeinside=||,mathescape=true]{python}{boundary_condition.fill_boundaries_levelset()}}\\
        \textcolor{blue}{\mintinline[escapeinside=||,mathescape=true]{python}{levelset_handler.compute_geometrical_quantities()}}\\
        \textcolor{blue}{\mintinline[escapeinside=||,mathescape=true]{python}{levelset_handler.mix_conservatives()}}\\
        \textcolor{blue}{\mintinline[escapeinside=||,mathescape=true]{python}{levelset_handler.transform_conservatives_to_volume_averages()}}\\
        \mintinline[escapeinside=||,mathescape=true]{python}{get_primitives_from_conservatives()}\\
        \textcolor{blue}{\mintinline[escapeinside=||,mathescape=true]{python}{levelset_handler.extend_primitives_into_ghost_cells()}}\\
        \mintinline[escapeinside=||,mathescape=true]{python}{boundary_condition.fill_material_boundaries()}\\
    }
\end{algorithm}

\begin{algorithm}
    \caption{Loop over spatial dimensions in the \mintinline[escapeinside=||,mathescape=true]{python}{compute_rhs} function.
    Blue text color indicates functions that are only executed for two-phase simulations.}
    \label{alg:compute_rhs}
    \For{active axis}{
        \mintinline[escapeinside=||,mathescape=true]{python}{flux_computer.compute_inviscid_flux_xi()}\\
        \mintinline[escapeinside=||,mathescape=true]{python}{flux_computer.compute_viscous_flux_xi()}\\
        \mintinline[escapeinside=||,mathescape=true]{python}{flux_computer.compute_heat_flux_xi()}\\
        \textcolor{blue}{\mintinline[escapeinside=||,mathescape=true]{python}{levelset_handler.weight_cell_face_flux_xi()}}\\
        \textcolor{blue}{\mintinline[escapeinside=||,mathescape=true]{python}{levelset_handler.compute_interface_flux_xi()}}\\
        \textcolor{blue}{\mintinline[escapeinside=||,mathescape=true]{python}{levelset_handler.compute_levelset_advection_rhs()}}\\
    }
\end{algorithm}

\subsection{Gradient Computation in JAX-FLUIDS}
\label{subsec:gradients}
JAX-FLUIDS offers the \mintinline{python}{simulate} method to perform a standard forward CFD simulation. In this regard, JAX-FLUIDS 
serves as a physics simulator for data generation, development of numerical methods, and exploration of fluid dynamics.
The \mintinline{python}{simulate} method does not have a return value, therefore, it is not meant to be used for
end-to-end optimization. 

To use the automatic differentiation capabilities of JAX-FLUIDS, we offer the \mintinline{python}{feed_forward} method.
This method takes in a batch of initial buffers and propagates them in time for a fixed number of time steps. The user provides 
the number of integration steps and a fixed time step. The output of the \mintinline{python}{feed_forward} method is the solution trajectory. 
In particular, the shape of the initial buffer is $(N_b, 5, N_x, N_y, N_z)$ where $N_b$ is the batch size. The shape of 
the solution trajectory is $(N_b, N_T + 1, 5, N_x, N_y, N_z)$ where $N_T$ is the number of integration steps. 
Internally, the \mintinline{python}{feed_forward} method uses the \mintinline{python}{jax.vmap} routine to vectorize 
over the batch dimension. \mintinline{python}{feed_forward} is jit-compilable and can be differentiated with
\mintinline{python}{jax.grad} or \mintinline{python}{jax.value_and_grad}. The \mintinline{python}{feed_forward} method 
of JAX-FLUIDS can therefore be used for end-to-end optimization of ML models.

\subsection{Integration of ML models into JAX-FLUIDS}
\label{subsec:MLJAXFLUIDS}
JAX-FLUIDS works with Haiku \cite{haiku2020github} and Optax \cite{optax2020github}.
The Haiku package is a neural network library for JAX.
In Haiku, neural networks are of type \mintinline{python}{haiku.Module}.
To use them in combination with JAX, the feedforward method of the network has to be embedded in a function
that is transformed into a forward wrapper object of type \mintinline{python}{haiku.Transformed}.
This forward wrapper object provides two pure methods, \mintinline{python}{init} and \mintinline{python}{apply}.
The \mintinline{python}{init} method initializes the network parameters, and 
the \mintinline{python}{apply} method executes the feedforward of the network.
Network parameters have to be explicitly passed to the \mintinline{python}{apply} method.
We refer to the Haiku documentation for more details.
Optax provides optimization routines, e.g., the popular Adam optimizer \cite{Kingma2015}.

In JAX-FLUIDS, we provide functionality to include preexisting ML models and optimize new ones.
Neural networks can be passed to the \mintinline{python}{simulate} and \mintinline{python}{feed_forward} method.
Note that only the \mintinline{python}{feed_forward} method can be differentiated, see Subsection \ref{subsec:gradients}.
I.e., it must be used for optimization of deep learning models.
A typical use case in ML-CFD research is substituting a conventional numerical subroutine with a data-driven alternative.
We provide a number of interfaces inside the JAX-FLUIDS solver to which network modules 
can be passed from \mintinline{python}{feed_forward}, e.g., to the cell face reconstruction, the Riemann solver,
or the forcing module.
On the top level, the user gives a dictionary of transformed network modules 
and a dictionary with corresponding network parameters to the \mintinline{python}{feed_forward} method.
The keys of these dictionaries specify the JAX-FLUIDS subroutine to which the corresponding values are passed.
\section{Validation of JAX-FLUIDS as a Classical CFD Simulator}
\label{sec:ForwardPass}
We show the capabilities of the JAX-FLUIDS solver as a classical fluid dynamics simulator.
We validate our implementation on established one- and two-dimensional test cases from gas dynamics and several canonical turbulent flows. 
In Subsection \ref{subsec:SinglePhase} we first validate the single-phase implementation in JAX-FLUIDS.
In Subsection \ref{subsec:Twophase} we then validate the two-phase level-set implementation for fluid-solid and fluid-fluid interactions.

We define two numerical setups which we will use predominantly throughout this section.
Firstly, we use the \textit{High-Order Godunov} formulation. This setup consists of a WENO5-JS reconstruction of the primitive variables
with an approximate HLLC Riemann solver.
We will refer to this setup as \textit{HLLC}.
Secondly, we employ the \textit{Flux-Splitting} formulation. In particular, we choose Roe approximate Riemann solver
with a WENO5-JS flux reconstruction in characteristic space.
This setup will be denoted as \textit{ROE}.
The dissipative fluxes are discretized with a fourth-order central finite difference stencil as described in \ref{subsec:Dissipative}.
If not specified otherwise we will use the stiffened gas equation of state with $\gamma = 1.4$ and $B = 0$, i.e., the ideal gas law. 
We will use the classical TVD-RK3 scheme with $CFL_\text{conservatives} = 0.9$ for time integration.
For the two-phase test cases, we additionally apply the following methods.
The level-set advection equation is discretized with a HOUC5 stencil.
We solve the extension equation using a first-order upwind spatial derivative stencil combined with an Euler integration scheme. 
Here, we apply a fixed number of $15$ steps with $CFL_\text{extension}=0.7$.
The reinitialization equation is solved with a WENO3-HJ stencil combined with a TVD-RK2 scheme. 
The level-set field is reinitialized each physical time step by integrating the reinitialization equation 
for one step with $CFL_\text{reinitialization}=0.7$.
We refer to Section \ref{sec:numericalmodel} and therein to Table \ref{tab:NumericalMethods} for details on numerical models.

\subsection{Single Phase Simulations}
\label{subsec:SinglePhase}

\subsubsection{Convergence Study}
\label{subsubsec:Convergence}

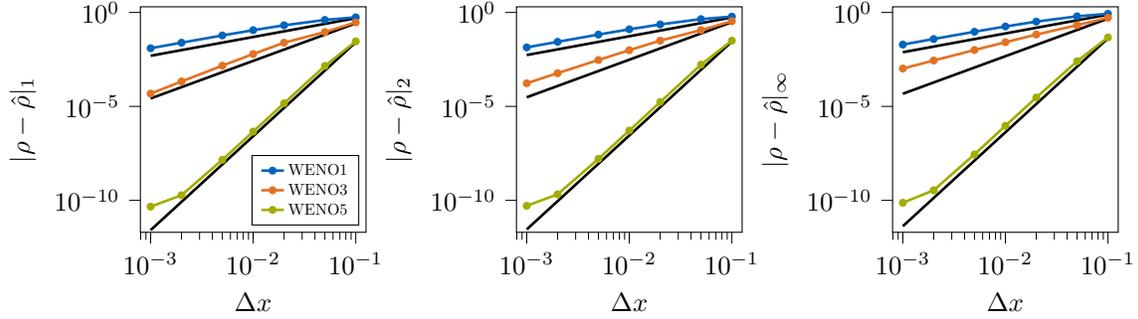
\begin{figure}
    \centering
\begin{tikzpicture}

\definecolor{color0}{RGB}{0,101,189}
\definecolor{color1}{RGB}{227,114,34}
\definecolor{color2}{RGB}{162,173,0}

\begin{groupplot}[group style={group size=3 by 1, horizontal sep=2.0cm}, height=3cm, width=3cm, scale only axis]
\nextgroupplot[
log basis x={10},
log basis y={10},
tick align=outside,
tick pos=left,
legend cell align={left},
legend style={
    at={(0.97,0.03)},
    anchor=south east,
    nodes={scale=0.6, transform shape}
},
xlabel={\(\displaystyle \Delta x\)},
xmin=0.000794328234724281, xmax=0.125892541179417,
xmode=log,
xtick style={color=black},
ylabel={\(\displaystyle \vert\rho - \hat{\rho} \vert_1\)},
ymin=2e-12, ymax=2,
ymode=log,
ytick style={color=black}
]
\addplot [line width=1, black, forget plot]
table {figures/fig_convergence/convergence-001.dat};
\addplot [line width=1, color0, mark=*, mark size=1, mark options={solid}]
table {figures/fig_convergence/convergence-000.dat};
\addlegendentry{WENO1};
\addplot [line width=1, black, forget plot]
table {figures/fig_convergence/convergence-003.dat};
\addplot [line width=1, color1, mark=*, mark size=1, mark options={solid}]
table {figures/fig_convergence/convergence-002.dat};
\addlegendentry{WENO3};
\addplot [line width=1, black, forget plot]
table {figures/fig_convergence/convergence-005.dat};
\addplot [line width=1, color2, mark=*, mark size=1, mark options={solid}]
table {figures/fig_convergence/convergence-004.dat};
\addlegendentry{WENO5};

\nextgroupplot[
log basis x={10},
log basis y={10},
tick align=outside,
tick pos=left,
xlabel={\(\displaystyle \Delta x\)},
xmin=0.000794328234724281, xmax=0.125892541179417,
xmode=log,
xtick style={color=black},
ylabel={\(\displaystyle \vert\rho - \hat{\rho} \vert_2\)},
ymin=2e-12, ymax=2,
ymode=log,
ytick style={color=black}
]
\addplot [line width=1, black]
table {figures/fig_convergence/convergence-007.dat};
\addplot [line width=1, color0, mark=*, mark size=1, mark options={solid}]
table {figures/fig_convergence/convergence-006.dat};
\addplot [line width=1, black]
table {figures/fig_convergence/convergence-009.dat};
\addplot [line width=1, color1, mark=*, mark size=1, mark options={solid}]
table {figures/fig_convergence/convergence-008.dat};
\addplot [line width=1, black]
table {figures/fig_convergence/convergence-011.dat};
\addplot [line width=1, color2, mark=*, mark size=1, mark options={solid}]
table {figures/fig_convergence/convergence-010.dat};

\nextgroupplot[
log basis x={10},
log basis y={10},
tick align=outside,
tick pos=left,
xlabel={\(\displaystyle \Delta x\)},
xmin=0.000794328234724281, xmax=0.125892541179417,
xmode=log,
xtick style={color=black},
ylabel={\(\displaystyle \vert\rho - \hat{\rho} \vert_\infty\)},
ymin=2e-12, ymax=2,
ymode=log,
ytick style={color=black}
]
\addplot [line width=1, black]
table {figures/fig_convergence/convergence-013.dat};
\addplot [line width=1, color0, mark=*, mark size=1, mark options={solid}]
table {figures/fig_convergence/convergence-012.dat};
\addplot [line width=1, black]
table {figures/fig_convergence/convergence-015.dat};
\addplot [line width=1, color1, mark=*, mark size=1, mark options={solid}]
table {figures/fig_convergence/convergence-014.dat};
\addplot [line width=1, black]
table {figures/fig_convergence/convergence-017.dat};
\addplot [line width=1, color2, mark=*, mark size=1, mark options={solid}]
table {figures/fig_convergence/convergence-016.dat};
\end{groupplot}

\end{tikzpicture}
    \caption{Error convergence $\vert \rho - \hat{\rho} \vert_p $ for the linear advection of Equation \eqref{eq:SinusDensity} 
    with WENO1, WENO3, and WENO5 spatial discretization.
    $\hat{\rho}$ is the analytical solution at $t_{\text{end}} = 1.0$.
    From left to right: $l_1$, $l_2$, and $l_{\infty}$ norms.}
    \label{fig:convergence}
\end{figure}

We analyze the convergence behavior of our solver.
We simulate the advection of a density profile in a one-dimensional periodic domain of extent $x \in [0, 1]$ with constant velocity $u=1$ and pressure $p=1$.
We use a sinusoidal initial condition for the density
\begin{equation}
    \rho(x, t=0) = 1.5 + \sin(2\pi x).
    \label{eq:SinusDensity}    
\end{equation}
Note that we initialize the cell-averaged values of Equation \eqref{eq:SinusDensity}.
I.e., for cell $i$ with cell-center $x_i$ we use $\bar{\rho}_i = 1.5 - 1.0 / (2 \pi \Delta x) (\cos(2  \pi x_{i+1/2}) - \cos(2 \pi x_{i-1/2})) $.
We conduct simulations with WENO1-JS, WENO3-JS, and WENO5-JS spatial discretizations and evaluate the rate of convergence upon mesh refinement.
We use TVD-RK2 for WENO1- and WENO3-JS. 
For WENO5-JS we use TVD-RK3.
We use a fixed time step $\Delta t = 1 \times 10^{-4}$ which is chosen small enough to exclude any influence of the time integration scheme.
The simulation is propagated until $t_{\text{end}} = 1.0$.
Figure \ref{fig:convergence} shows the convergence rates in the $l_1$, $l_2$, and $l_\infty$ norms.
We increase the resolution from $10$ to $1000$ points.
The expected convergence rates of $\mathcal{O}(\Delta x^1)$, $\mathcal{O}(\Delta x^2)$, and $\mathcal{O}(\Delta x^5)$ are reached.
Note that it is well known that the convergence order of WENO3-JS drops to second order under the presence of extreme points.  

\subsubsection{Shock Tube Tests}
\label{subsubsec:Shocktube}
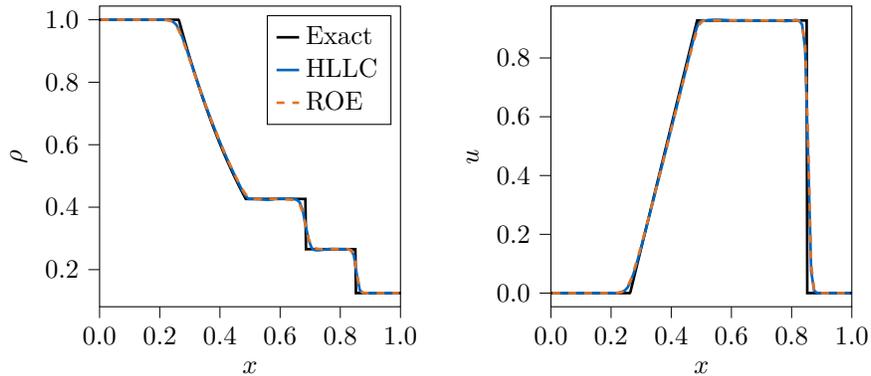
\begin{figure}
    \centering
\begin{tikzpicture}

\definecolor{color0}{RGB}{0,101,189}
\definecolor{color1}{RGB}{227,114,34}
\definecolor{color2}{RGB}{162,173,0}

\begin{groupplot}[group style={group size=2 by 1, horizontal sep=2.0cm}, height=4cm, width=4cm, scale only axis]
\nextgroupplot[
legend cell align={left},
legend style={anchor=north east, at={(0.97, 0.97)}},
tick align=outside,
tick pos=left,
xlabel={\(\displaystyle x\)},
xtick={0,0.2,0.4,0.6,0.8,1},
xticklabels={0.0,0.2,0.4,0.6,0.8,1.0},
ytick={0,0.2,0.4,0.6,0.8,1},
yticklabels={0.0,0.2,0.4,0.6,0.8,1.0},
xmin=0, xmax=1,
xtick style={color=black},
ylabel={\(\displaystyle \rho\)},
ymin=0.0812499996904965, ymax=1.0437500006658,
ytick style={color=black}
]
\addplot [line width=1,black]
table {figures/fig_shocktube_1/shocktube_1-002.dat};
\addlegendentry{Exact}
\addplot [line width=1, color0]
table {figures/fig_shocktube_1/shocktube_1-000.dat};
\addlegendentry{HLLC}
\addplot [line width=1, color1, dashed]
table {figures/fig_shocktube_1/shocktube_1-001.dat};
\addlegendentry{ROE}

\nextgroupplot[
tick align=outside,
tick pos=left,
xlabel={\(\displaystyle x\)},
xmin=0, xmax=1,
xtick style={color=black},
ylabel={\(\displaystyle u\)},
ymin=-0.0465018608028844, ymax=0.976539027476759,
ytick style={color=black},
xtick={0,0.2,0.4,0.6,0.8,1},
xticklabels={0.0,0.2,0.4,0.6,0.8,1.0},
ytick={0,0.2,0.4,0.6,0.8,1},
yticklabels={0.0,0.2,0.4,0.6,0.8,1.0},
]
\addplot [line width=1, black]
table {figures/fig_shocktube_1/shocktube_1-005.dat};
\addplot [line width=1, color0]
table {figures/fig_shocktube_1/shocktube_1-003.dat};
\addplot [line width=1, color1, dashed]
table {figures/fig_shocktube_1/shocktube_1-004.dat};

\end{groupplot}

\end{tikzpicture}
    \caption{Sod shock tube: density $\rho$ and velocity $u$ at $t=0.2$.}
    \label{fig:shocktube_1}
\end{figure}

\begin{figure}
    \centering
\begin{tikzpicture}

\definecolor{color0}{RGB}{0,101,189}
\definecolor{color1}{RGB}{227,114,34}
\definecolor{color2}{RGB}{162,173,0}

\begin{groupplot}[group style={group size=2 by 1, horizontal sep=2.0cm}, height=4cm, width=4cm, scale only axis]
\nextgroupplot[
legend cell align={left},
legend style={anchor=north west, at={(0.03, 0.97)}},
tick align=outside,
tick pos=left,
xlabel={\(\displaystyle x\)},
xmin=0, xmax=1,
xtick style={color=black},
ylabel={\(\displaystyle \rho\)},
ymin=0.290310220180997, ymax=1.35332065575171,
ytick style={color=black},
xtick={0,0.2,0.4,0.6,0.8,1},
xticklabels={0.0,0.2,0.4,0.6,0.8,1.0},
ytick={0.4,0.8,1.2},
yticklabels={0.4,0.8,1.2},
]
\addplot [line width=1, black]
table {figures/fig_shocktube_2/shocktube_2-002.dat};
\addlegendentry{Exact}
\addplot [line width=1, color0]
table {figures/fig_shocktube_2/shocktube_2-000.dat};
\addlegendentry{HLLC}
\addplot [line width=1, color1, dashed]
table {figures/fig_shocktube_2/shocktube_2-001.dat};
\addlegendentry{ROE}

\nextgroupplot[
tick align=outside,
tick pos=left,
xlabel={\(\displaystyle x\)},
xmin=0, xmax=1,
xtick style={color=black},
ylabel={\(\displaystyle u\)},
ymin=-0.0789104600421587, ymax=1.65,
ytick style={color=black},
xtick={0,0.2,0.4,0.6,0.8,1},
xticklabels={0.0,0.2,0.4,0.6,0.8,1.0},
ytick={0,0.5,1,1.5},
yticklabels={0.0,0.5,1.0,1.5},
]
\addplot [line width=1, black]
table {figures/fig_shocktube_2/shocktube_2-005.dat};
\addplot [line width=1, color0]
table {figures/fig_shocktube_2/shocktube_2-003.dat};
\addplot [line width=1, color1, dashed]
table {figures/fig_shocktube_2/shocktube_2-004.dat};
\end{groupplot}

\end{tikzpicture}
    \caption{Lax shock tube: density $\rho$ and velocity $u$ at $t=0.14$.}
    \label{fig:shocktube_2}
\end{figure}
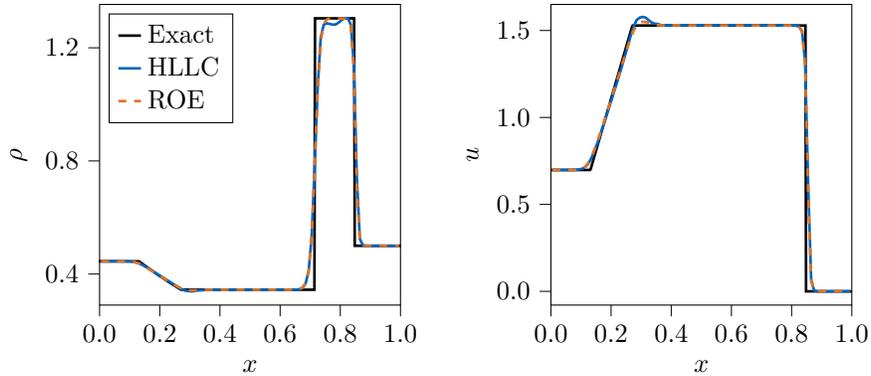

The shock tube tests of Sod \cite{Sod1978a} and Lax \cite{Lax1954} are standard one-dimensional test cases for validating
fluid solvers for compressible flows.
Specifically, we use these test cases to validate the implementation of the convective fluxes and the shock-capturing schemes.
In both shock tube test cases three waves emanate from the initial discontinuities:
a left running rarefaction, a right running contact discontinuity, and a right running shock.
A detailed description of the tests and their setups are provided in the presented references.
We discretize the domain $x \in [0, 1]$ with $N = 100$ points.
The analytical reference solution is taken from an exact Riemann solver (e.g, \cite{Toro2009a}).
We run both shock tube tests with the \textit{HLLC} and \textit{ROE} setups.
Figures \ref{fig:shocktube_1} and \ref{fig:shocktube_2} show the density and velocity distributions for the Sod and Lax shock tube tests at $t=0.2$ and $t=0.14$, respectively.
The \textit{HLLC} and \textit{ROE} solutions agree very well with the analytical reference.
The \textit{HLLC} solutions show slightly oscillatory behavior which is due to the cell face reconstruction of the primitive variables, see \cite{Qiu2002}. 

\subsubsection{Lid-driven Cavity}
\label{subsubsec:Cavity}
\begin{figure}
    \centering
    \input{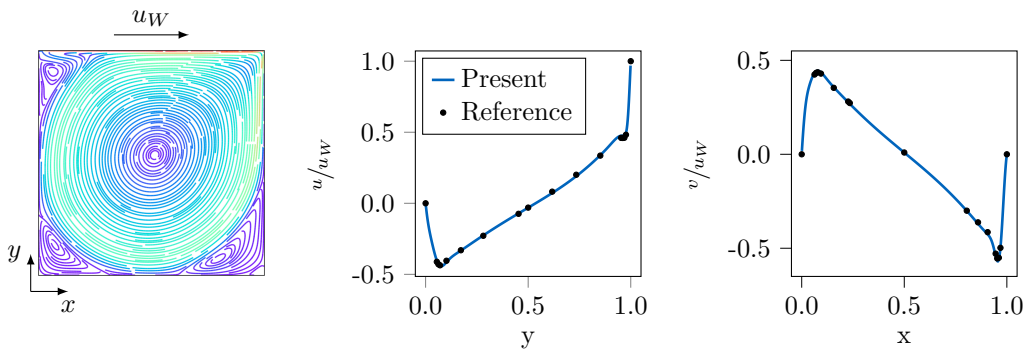}
    \caption{Lid driven cavity at $Re = \sfrac{u_W L}{\nu} = 5000$. (Left) Instantaneous streamlines with colors ranging 
    from largest (red) and smallest (blue) value of the absolute velocity.
    (Middle) Normalized $u$ velocity over $y$ through domain center.
    (Right) Normalized $v$ velocity over $x$ through domain center. Reference is taken from \cite{Ghia1982}.}
    \label{fig:driven_cavity}
\end{figure}

The lid-driven cavity test case describes the flow in a square cavity that is driven by a moving wall.
Viscous forces lead to the generation of one primary vortex and, depending on the Reynolds number, one or more 
secondary vortices. We use this test case to validate the implementation of the viscous fluxes. 
The computational domain $x\times y\in[0,1]\times[0,1]$ is discretized with a grid consisting of $500\times500$ cells. 
All boundary conditions are no-slip walls. 
The north wall is moving with a constant velocity $u_W$, resulting in a Reynolds number $Re=\frac{u_W L}{\nu} = 5000$. 
The test case is simulated until a steady state is reached with the \textit{HLLC} setup.
Figure \ref{fig:driven_cavity} depicts the distribution of $\frac{u}{u_W}$ over $y$ and $\frac{v}{u_W}$ over $x$ across the domain center. 
The present result agrees very well with the reference \cite{Ghia1982}.

\subsubsection{Compressible Decaying Isotropic Turbulence}
\label{subsubsec:HIT}
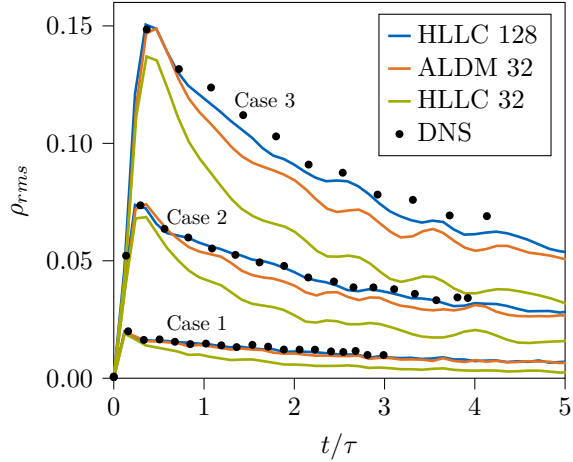
\begin{figure}
    \centering
\begin{tikzpicture}

\definecolor{color0}{RGB}{0,101,189}
\definecolor{color1}{RGB}{227,114,34}
\definecolor{color2}{RGB}{162,173,0}

\begin{axis}[
scale only axis,
height=5cm, width=6cm,
legend cell align={left},
legend style={at={(0.97,0.97)}, anchor=north east},
tick align=outside,
tick pos=left,
xlabel={\(\displaystyle t / \tau\)},
xmin=0, xmax=5,
xtick style={color=black},
ylabel={\(\displaystyle \rho_{rms}\)},
ymin=0.0, ymax=0.16,
ytick style={color=black},
ytick = {0.0,0.05,0.1,0.15},
yticklabels = {0.00,0.05,0.10,0.15},
]
\addplot [line width=1,forget plot, color0]
table {figures/fig_decaying_turbulence/density_rms_decay-000.dat};
\addplot [line width=1,forget plot, color0]
table {figures/fig_decaying_turbulence/density_rms_decay-001.dat};
\addplot [line width=1, color0]
table {figures/fig_decaying_turbulence/density_rms_decay-002.dat};
\addlegendentry{HLLC 128}
\addplot [line width=1,forget plot, color1]
table {figures/fig_decaying_turbulence/density_rms_decay-003.dat};
\addplot [line width=1,forget plot, color1]
table {figures/fig_decaying_turbulence/density_rms_decay-004.dat};
\addplot [line width=1, color1]
table {figures/fig_decaying_turbulence/density_rms_decay-005.dat};
\addlegendentry{ALDM 32}
\addplot [line width=1,forget plot, color2]
table {figures/fig_decaying_turbulence/density_rms_decay-006.dat};
\addplot [line width=1,forget plot, color2]
table {figures/fig_decaying_turbulence/density_rms_decay-007.dat};
\addplot [line width=1, color2]
table {figures/fig_decaying_turbulence/density_rms_decay-008.dat};
\addlegendentry{HLLC 32}
\addplot [line width=1,forget plot, black, mark=*, mark size=1, mark options={solid,fill=black}, only marks, forget plot]
table {figures/fig_decaying_turbulence/density_rms_decay-009.dat};
\addplot [line width=1,forget plot, black, mark=*, mark size=1, mark options={solid,fill=black}, only marks, forget plot]
table {figures/fig_decaying_turbulence/density_rms_decay-010.dat};
\addplot [line width=1, black, mark=*, mark size=1, mark options={solid,fill=black}, only marks]
table {figures/fig_decaying_turbulence/density_rms_decay-011.dat};
\addlegendentry{DNS}
\draw (axis cs:0.5,0.03) node[
  scale=0.8,
  anchor=north west,
  text=black,
  rotate=0.0
]{Case 1};
\draw (axis cs:0.5,0.075) node[
  scale=0.8,
  anchor=north west,
  text=black,
  rotate=0.0
]{Case 2};
\draw (axis cs:1.25,0.125) node[
  scale=0.8,
  anchor=north west,
  text=black,
  rotate=0.0
]{Case 3};
\end{axis}

\end{tikzpicture}
    \caption{Temporal evolution of the rms density fluctuations for decaying isotropic turbulence.
    The time axis is normalized with the initial eddy turnover time.
    The DNS results are cases 1-3 from Spyropoulos et al. \cite{Spyropoulos1996}.}
    \label{fig:densityrmsdecay}
\end{figure}

Many flows in nature and engineering applications are turbulent.
The direct numerical simulation of turbulent flows requires the resolution of the smallest scales present (the Kolmogorov scales) which is prohibitively expansive \cite{Pope2000}.
In JAX-FLUIDS we have implemented the implicit large eddy simulation (ILES) model \textit{ALDM} \cite{Hickel2014b} which enables us to simulate complex compressible turbulent flows
up to very high Reynolds numbers.
To validate the implementation of \textit{ALDM} and the viscous forces, we simulate compressible decaying isotropic turbulence at various turbulent Mach numbers.
Specifically, we investigate the performance of the \textit{ALDM} ILES implementation on the basis of cases 1-3 from Spyropoulos et al. \cite{Spyropoulos1996}.
The turbulent Mach numbers $M_t = \sqrt{<u^2 + v^2 + w^2>} / <c>$ for the three cases are $0.2$, $0.4$, and $0.6$, respectively.
The turbulent Reynolds number is $Re_T = \frac{\rho u'^4}{\epsilon \mu} = 2742$,
where $u'$ is the root-mean-square (rms) velocity and $\epsilon$ is the dissipation rate of turbulent kinetic energy \cite{Pope2000}.
The spatial domain has extent $x \times y \times z \in \left[0, 2\pi\right] \times \left[0, 2\pi\right] \times \left[0, 2\pi\right]$.
We use the DNS data from \cite{Spyropoulos1996} and an \textit{HLLC} simulation with a resolution of $128^3$ cells as reference.
The LES simulations are performed on a coarse grid with $32^3$ cells.
We use \textit{ALDM} and \textit{HLLC} on this resolution to conduct LES simulations.
The initial velocity field is divergence free and has the energy spectrum
\begin{align}
    E(k) = A k^4 \exp(-2 k^2 / k_0^2),
\end{align}
where $k$ is the wave number, $k_0$ is the wave number at which the spectrum is maximal, and $A$ is a constant chosen to adjust a specified initial kinetic energy.
The initial density and temperature fields are uniform.
Figure \ref{fig:densityrmsdecay} shows the temporal evolution of the rms density fluctuations $\rho_{rms} = \sqrt{<\rho' \rho'>}$.
We normalize the time axis with the initial eddy turnover time $\tau = \lambda_f / u' \approx 0.85$.
$\lambda_f$ is the lateral Taylor microscale \cite{Pope2000}.
The \textit{HLLC} simulation at $128^3$ recovers the DNS reference very well.
On the coarse mesh, the performance of the \textit{ALDM} ILES becomes obvious when compared to \textit{HLLC} simulations at the same resolution.
\textit{ALDM} gives good results consistent with the DNS data, indicating that \textit{ALDM} recovers the correct sub-grid scale terms for compressible isotropic turbulence.

\subsection{Two-phase Simulations}
\label{subsec:Twophase}

\subsubsection{Two-phase Sod Shock Tube}
\label{subsubsec:TwophaseSod}
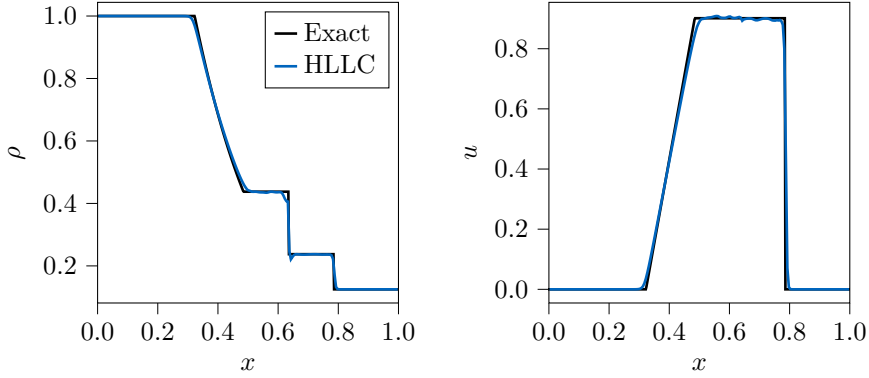
\begin{figure}
    \centering
\begin{tikzpicture}

\definecolor{color0}{RGB}{0,101,189}
\definecolor{color1}{RGB}{227,114,34}
\definecolor{color2}{RGB}{162,173,0}

\begin{groupplot}[group style={group size=2 by 1, horizontal sep=2.0cm}, height=4cm, width=4cm, scale only axis]
\nextgroupplot[
legend cell align={left},
legend style={anchor=north east, at={(0.97, 0.97)}},
tick align=outside,
tick pos=left,
xlabel={\(\displaystyle x\)},
xmin=0, xmax=1,
xtick style={color=black},
ylabel={\(\displaystyle \rho\)},
ymin=0.0812499999089802, ymax=1.04375000074439,
ytick style={color=black},
xtick={0,0.2,0.4,0.6,0.8,1},
xticklabels={0.0,0.2,0.4,0.6,0.8,1.0},
ytick={0,0.2,0.4,0.6,0.8,1},
yticklabels={0.0,0.2,0.4,0.6,0.8,1.0},
]
\addplot [line width=1, black]
table {figures/fig_twophase_shocktube/twophase_shocktube-000.dat};
\addlegendentry{Exact}
\addplot [line width=1, color0]
table {figures/fig_twophase_shocktube/twophase_shocktube-001.dat};
\addlegendentry{HLLC}

\nextgroupplot[
tick align=outside,
tick pos=left,
xlabel={\(\displaystyle x\)},
xmin=0, xmax=1,
xtick style={color=black},
ylabel={\(\displaystyle u\)},
ymin=-0.0454387018610033, ymax=0.954212720729219,
ytick style={color=black},
xtick={0,0.2,0.4,0.6,0.8,1},
xticklabels={0.0,0.2,0.4,0.6,0.8,1.0},
ytick={0,0.2,0.4,0.6,0.8,1},
yticklabels={0.0,0.2,0.4,0.6,0.8,1.0},
]
\addplot [line width=1, black]
table {figures/fig_twophase_shocktube/twophase_shocktube-002.dat};
\addplot [line width=1, color0]
table {figures/fig_twophase_shocktube/twophase_shocktube-003.dat};
\end{groupplot}

\end{tikzpicture}
    \caption{Air-helium shock tube problem.
    Density $\rho$ and velocity $u$ at $t = 0.15$.}
    \label{fig:twophaseshocktube}
\end{figure}

We consider a two-phase variant of the Sod shock tube problem \cite{Sod1978a}.
This test case validates the inviscid fluid-fluid interface interactions.
In particular, we investigate an air-helium shock tube problem 
in which the materials left and right of the initial discontinuity are air ($\gamma_{\text{air}} = 1.4$) 
and helium ($\gamma_{\text{helium}} = 1.667$), respectively.
We use the previously described \textit{HLLC} setup.
The domain $x \in [0, 1]$ is resolved with 200 cells.
Figure \ref{fig:twophaseshocktube} shows the results at $t = 0.15$.
The numerical approximations are in good agreement with the analytical solution.
The interface position and shock speed and shock strength are captured correctly.
We observe slight density oscillations around the interface.
This is in agreement with previous literature \cite{Hu2006} as no isobaric fix is employed.

\subsubsection{Bow Shock}
\label{subsubsec:Bowshock}
\begin{figure}
    \centering
    \begin{tikzpicture}
        \node (A) at (0,0) {\includegraphics[scale=0.4]{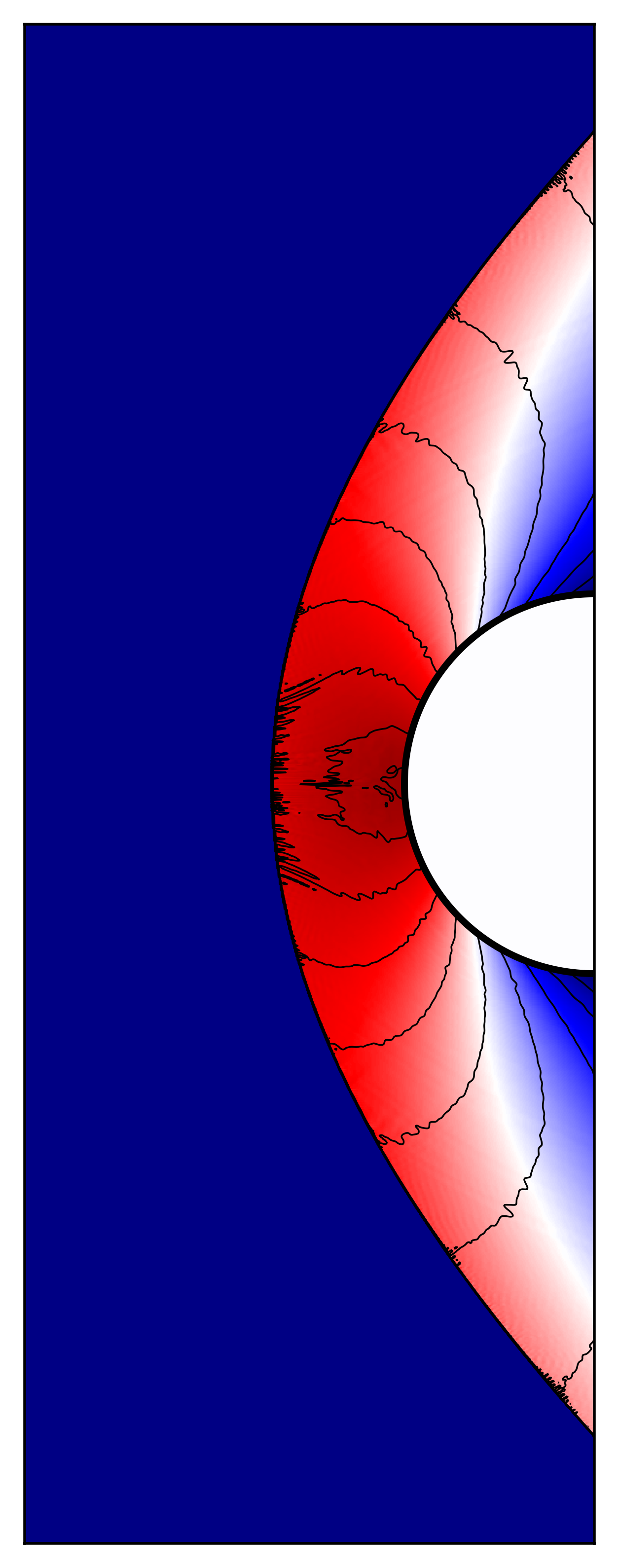}};
        \node [right = 0cm of A] (B) {\includegraphics[scale=0.4]{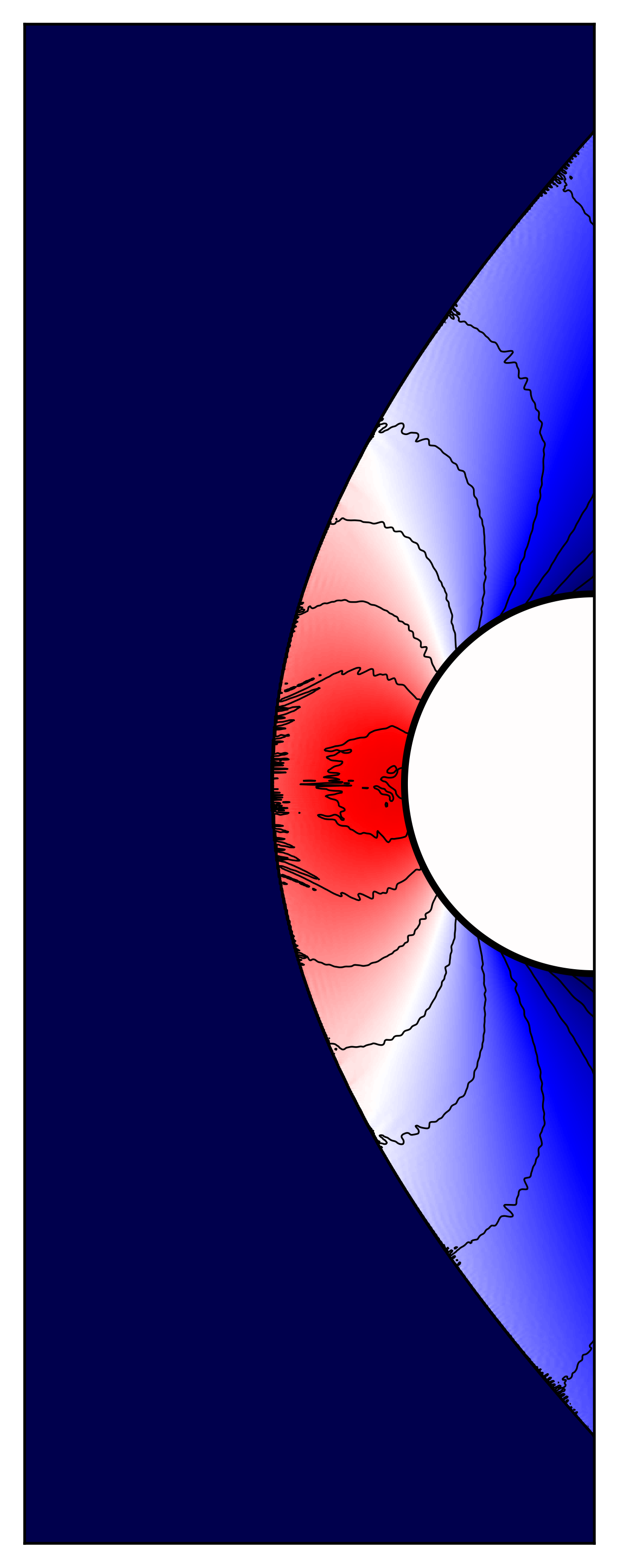}};
        \node at (A.north) {$\rho$};
        \node at (B.north) {$p$};
        \draw[->] ([xshift=-.1cm, yshift=-.2cm]A.south west) -- ([yshift=-.2cm, xshift=.4cm]A.south west) node[at end, below] {$x$};
        \draw[->] ([xshift=-.1cm, yshift=-.2cm]A.south west) -- ([yshift=.3cm, xshift=-.1cm]A.south west) node[at end, left] {$y$};

        \node [right = 2cm of B] (C)  {\includegraphics[scale=0.4]{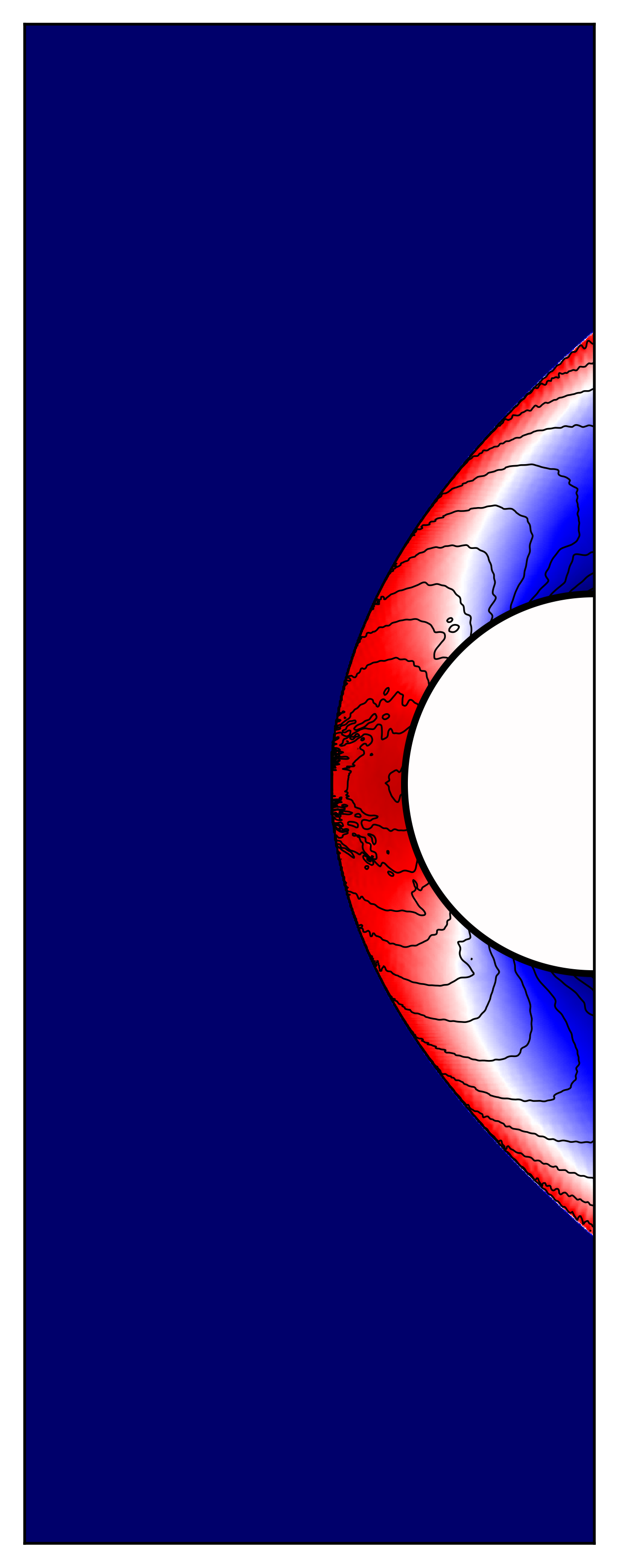}};
        \node [right = 0cm of C] (D) {\includegraphics[scale=0.4]{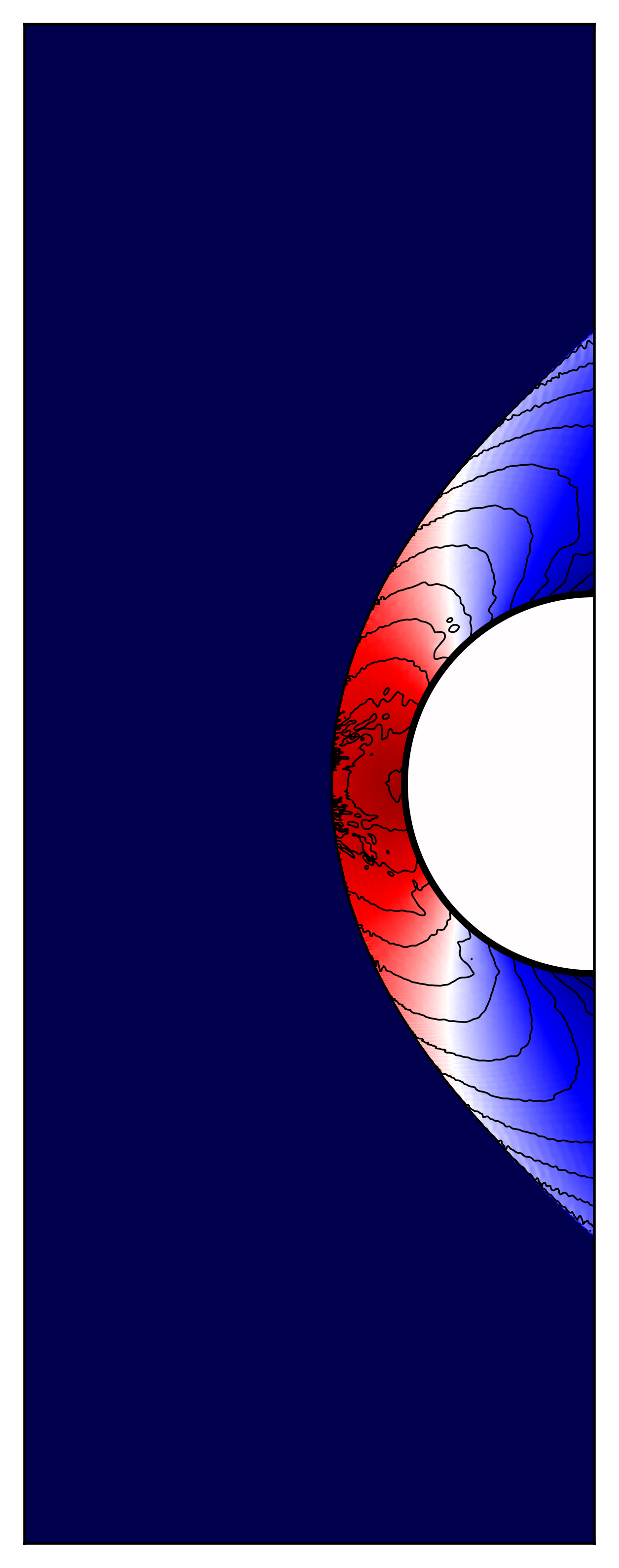}};
        \node at (C.north) {$\rho$};
        \node at (D.north) {$p$};
    \end{tikzpicture}
    \caption{Density and pressure for the bow shock at $Ma=3$ (left) and $Ma=20$ (right).
    The colormap ranges from minimum (blue) to maximum (red) value; $\rho\in[0.7, 4.5]$, $p\in[1.0, 15.0]$ for $Ma=3$
    and $\rho\in[0.7, 7.2]$, $p\in[1.0, 550.0]$ for $Ma=20$. The black lines represent Mach isocontours from 0.1 to 2.5 in steps of 0.2.}
    \label{fig:bowshock}
\end{figure}
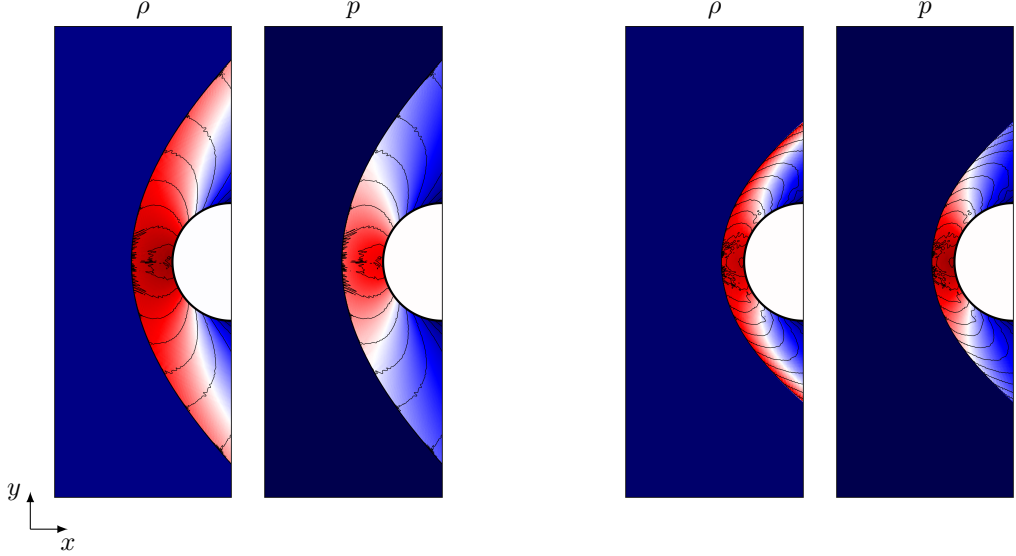

Bow shocks occur in supersonic flows around blunt bodies \cite{PEERY1988}. Here, we simulate the flow around a stationary cylinder at high 
Mach numbers $Ma = \sfrac{\sqrt{\mathbf{u} \cdot \mathbf{u}}}{\sqrt{\gamma\frac{p}{\rho}}}=\{3,20\}$. 
This test case validates the implementation of the inviscid fluid-solid
interface fluxes. The computational domain $x\times y\in[-0.3,0.0]\times[-0.4,0.4]$ is discretized with a grid consisting 
of $480\times1280$ cells. 
A cylinder with diameter $0.2$ is placed at the center of the east boundary.
The north, east, and south boundaries are zero-gradient. 
The west boundary is of Dirichlet type, imposing the post shock fluid state.
The fluid is initialized with the post shock state, i.e., $(\rho,u,v,p)=\left(1,\sqrt{1.4} Ma,0,1\right)$.
We simulate the test case with the \textit{HLLC} setup until a steady state solution is reached.
Figure \ref{fig:bowshock} illustrates the steady state density and pressure distributions. The results compare well to results from 
literature, e.g., \cite{Fleischmann2020}.

\subsubsection{Oscillating Drop}
\label{subsubsec:OscillatingDrop}
\begin{figure}[H]
    \centering
    \input{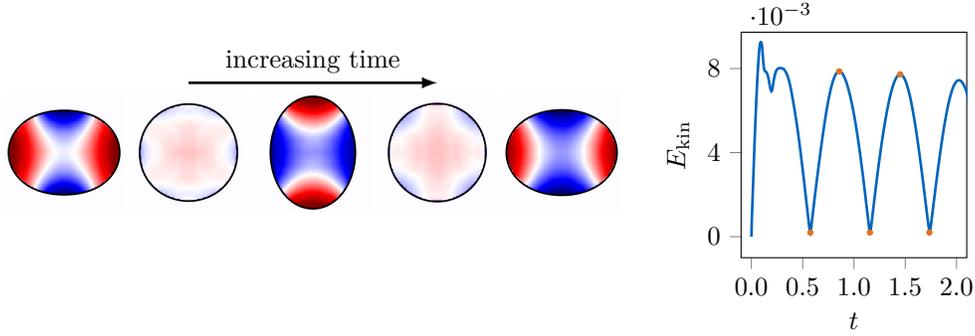}
    \caption{Oscillating drop. (Left) Temporal evolution of the pressure distribution within the drop. The colors range from maximum (red)
    to minimum (blue) value within the shown time period. (Right) Kinetic energy $E_\text{kin}=\int_V \rho \mathbf{u}\cdot\mathbf{u}\text{d}V$ of
    the drop over time. The orange dots indicate the times that correspond to the pressure distributions on the left.}
    \label{fig:oscillatingdrop}
\end{figure}

We consider a drop oscillating due to the interplay of surface tension and inertia.
Starting from an ellipsoidal shape, surface tension forces drive the drop to a circular shape. 
This process is associated with a transfer of potential to kinetic energy. 
The oscillating drop test case validates the implementation of the surface tension forces. 
The drop oscillates with a distinct frequency. 
The oscillation period $T$ is given by \cite{Rayleigh1879}
\begin{equation}
    \omega^2 = \frac{6\sigma}{(\rho_b+\rho_d)R^3}, \qquad T = \frac{2\pi}{\omega}.
\end{equation}
We discretize the computational domain $x\times y\in[0,1]\times[0,1]$ with a grid consisting of $200\times200$ cells.
We place an ellipse with semi-major and semi-minor axes of $0.2$ and $0.15$ at the center of the domain. 
The effective circle radius is therefore $R=0.17321$. 
The bulk and drop densities are $\rho_b=\rho_d=1.0$ and the surface tension coefficient is $\sigma=0.05$. 
All boundaries are zero-gradient. 
We use the \textit{HLLC} setup for this simulation. 
Figure \ref{fig:oscillatingdrop} displays instantaneous pressure fields and the kinetic energy of the drop over time.
The present result for the oscillation period is $T=1.16336$, which is in very good agreement with the analytical reference $T_{ref} = 1.16943$.

\subsubsection{Shear Drop Deformation}
The shear drop deformation test case describes the deformation of an initially circular shaped drop due to homogenous shear. 
Viscous forces lead to the deformation to an ellipsoidal shape. For stable parameters, surface tension forces will eventually
balance the viscous forces, which results in a steady state solution.
The shear flow is generated with moving no-slip wall boundaries. 
We use this test case to validate the viscous fluid-fluid interface fluxes. The steady state solution is
characterized by the viscosity ratio $\mu_b/\mu_d$ and the capillary number $Ca$
\begin{equation}
    Ca = \frac{\mu_b R \dot{s}}{\sigma},
\end{equation}
where $R$ denotes the initial drop radius, $\sigma$ the surface tension coefficient, $\dot{s}$ the shear rate, and 
$\mu_b$ and $\mu_d$ the viscosities of the drop and bulk fluid, respectively. 
The following relation holds for small deformations \cite{TaylorG.1934}.
\begin{equation}
    \qquad D = \frac{B_1 - B_2}{B_1 + B_2} = Ca \frac{19\mu_b/\mu_d+16}{16\mu_b/\mu_d+16}
\end{equation}
Herein, $B_1$ and $B_2$ indicate the semi-major and semi-minor axes of the steady state ellipse. 
To simulate this test case, we discretize the domain $x\times y\in[0,1]\times[0,1]$ with a grid that consists of $250\times250$ cells. 
A drop with radius $R=0.2$ is placed at the center of the domain. 
We move the north and south wall boundaries with an absolute velocity of $u_W = 0.1$ in positive and negative direction, respectively. 
This results in a shear rate of $0.2$. At the east and west boundaries we enforce periodicity. 
The viscosities are $\mu_b=\mu_d=0.1$. We simulate multiple capillary numbers by varying the surface tension coefficient with the \textit{HLLC} setup until a steady state solution is reached.
\label{subsubsec:ShearDrop}
\begin{figure}[H]
    \centering
    \input{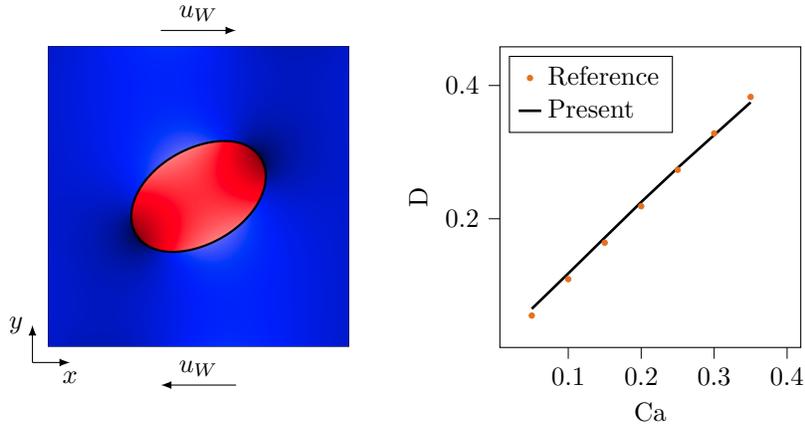}
    \caption{Shear drop deformation. (Left) Steady state pressure field for $Ca=0.2$. (Right) Deformation parameter $D$ over $Ca$.}
    \label{fig:sheardrop}
\end{figure}
Figure \ref{fig:sheardrop} illustrates the pressure distribution of the steady state ellipse at $Ca=0.2$. Furthermore, it shows the
deformation parameter $D$ over the capillary number $Ca$. The present result agrees well with the analytical reference.

\subsubsection{Shock-bubble Interaction}
\label{subsubsec:ShockBubble}

We simulate the interaction of a shock with Mach number 1.22 with a helium bubble immersed in air.
This is an established test case to assess the robustness and validity of compressible two-phase numerical methods.
Reports are well documented in literature \cite{Fedkiw1999a,Terashima2009,Hoppe2022a} and experimental data are available \cite{Haas1987}.

\begin{figure}
    \centering
    \begin{tikzpicture}
        \node (A) at (0,0) {\includegraphics[scale=0.6]{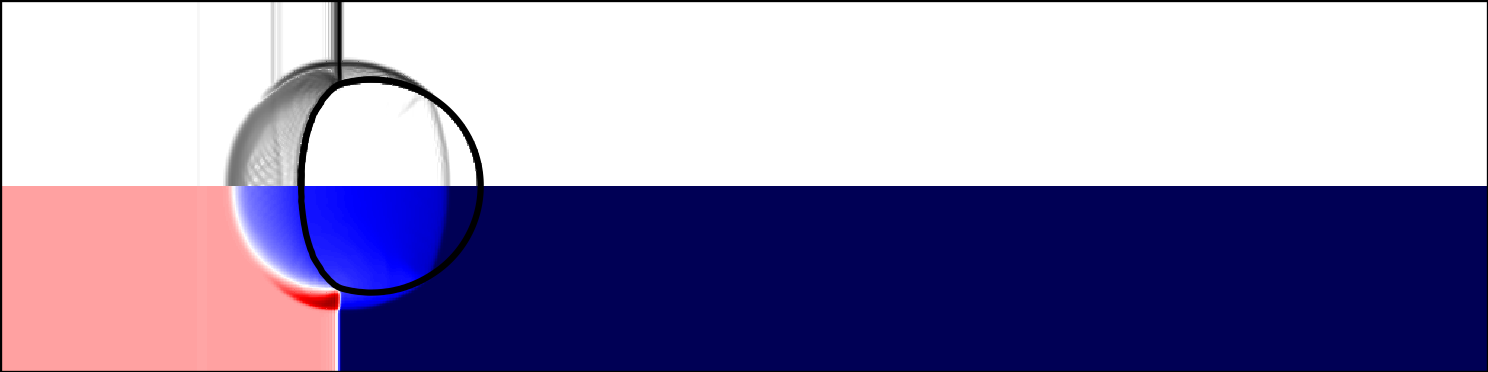}};
        \node at ([yshift=-.2cm]A.south) {$t = 55 \mu s$};

        \node [right = 0.0cm of A] (B)  {\includegraphics[scale=0.6]{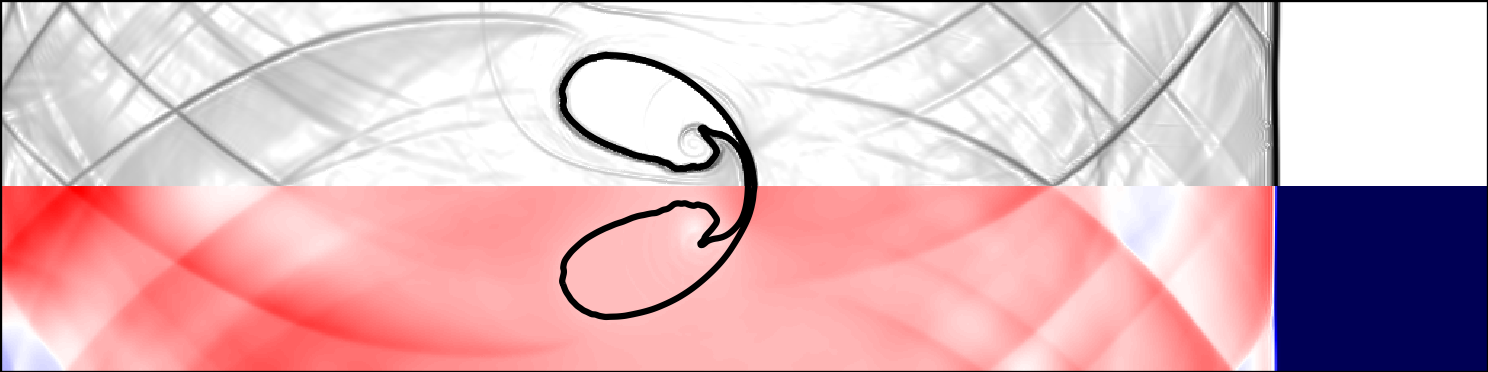}};
        \node at ([yshift=-.2cm]B.south) {$t = 555 \mu s$};
    \end{tikzpicture}
    \caption{Visualizations of the flow field at two different instances in time.
    The upper half of each image shows numerical schlieren, the lower half shows the pressure field
    from the smallest (blue) to the largest (red) pressure value.
    The black line indicates the location of the interface. 
    }
    \label{fig:shockbubbleflowfield}
\end{figure}

A helium bubble with diameter $D = 50\, \text{mm}$ is placed at the origin of the computational domain 
$x \times y \in [-90\, \text{mm}, 266\, \text{mm}] \times [-44.5\, \text{mm}, 44.5\, \text{mm}]$.
The initial shock wave is located $5\, \text{mm}$ to the left of the helium bubble.
The shock wave travels right and impacts with the helium bubble.
Figure \ref{fig:shockbubbleflowfield} shows the flow field at two later time instances.
The interaction of the initial shock with the helium bubble generates a reflected shock which is travelling to the left
and a second shock which is transmitted through the bubble, see Figure \ref{fig:shockbubbleflowfield} on the left.
The incident shock wave is visible as a vertical line.
The transmitted wave travels faster than the incident shock.
The helium bubbles deforms strongly and a re-entrant jet forms.
Figure \ref{fig:shockbubbleflowfield} on the right shows the instance in time at which the jet impinges on the interface of the bubble.

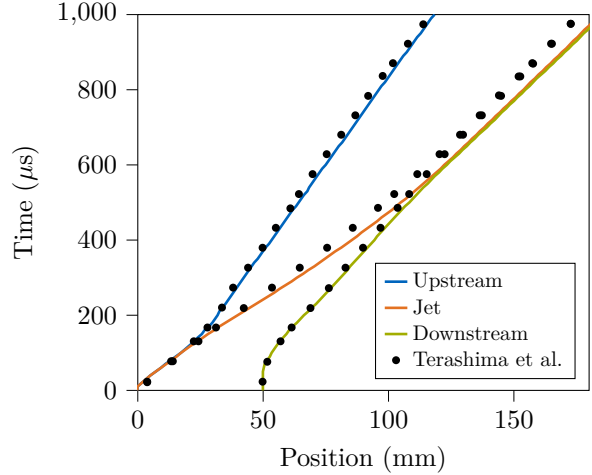
\begin{figure}
    \centering
\begin{tikzpicture}

\definecolor{color0}{RGB}{0,101,189}
\definecolor{color1}{RGB}{227,114,34}
\definecolor{color2}{RGB}{162,173,0}

\begin{axis}[
scale only axis,
width=6cm, height=5cm,
legend cell align={left},
legend style={
  at={(0.97,0.03)},
  anchor=south east,
  nodes={scale=0.8, transform shape}
},
tick align=outside,
tick pos=left,
xlabel={Position (mm)},
xmin=0, xmax=180,
xtick style={color=black},
ylabel={Time (\(\displaystyle \mu\)s)},
ymin=0, ymax=1000,
ytick style={color=black}
]
\addplot [line width=1, color0]
table {figures/fig_shockbubble/points-000.dat};
\addlegendentry{Upstream}
\addplot [line width=1, color1]
table {figures/fig_shockbubble/points-001.dat};
\addlegendentry{Jet}
\addplot [line width=1, color2]
table {figures/fig_shockbubble/points-002.dat};
\addlegendentry{Downstream}
\addplot [line width=1, black, mark=*, mark size=1, mark options={solid,fill=black,draw=black}, only marks]
table {figures/fig_shockbubble/points-003.dat};
\addlegendentry{Terashima et al.}
\addplot [line width=1, black, mark=*, mark size=1, mark options={solid,fill=black,draw=black}, only marks, forget plot]
table {figures/fig_shockbubble/points-004.dat};
\addplot [line width=1, black, mark=*, mark size=1, mark options={solid,fill=black,draw=black}, only marks, forget plot]
table {figures/fig_shockbubble/points-005.dat};
\end{axis}

\end{tikzpicture}
    \caption{Space-time diagram of three characteristic interface points.
    Positions of the upstream point (left-most point of the interface), the downstream point (right-most point of the interface),
    and the jet (left-most point of the interface on the center-line) are tracked.
    Reference values are taken from Terashima et al. \cite{Terashima2009}.}
    \label{fig:shockbubblespacetime}
\end{figure}

The numerical schlieren images and the flow fields in Figure \ref{fig:shockbubbleflowfield} are in good qualitative
agreement with results from literature, compare with Figure 7 from \cite{Haas1987} or Figure 10 from \cite{Hoppe2022a}.
Figure \ref{fig:shockbubblespacetime} shows the temporal evolution of characteristic interface points.
The results are in very good quantitative agreement with \cite{Terashima2009}.

\section{Single Node Performance}
\label{sec:Performance}
We assess the single node performance of the JAX-FLUIDS solver on an NVIDIA RTX A6000 GPU.
The NVIDIA RTX A6000 provides 48GB of GPU memory and a bandwidth of 768 GB/s. 
We conduct simulations of the three-dimensional compressible Taylor-Green vortex (TGV) \cite{Brachet1984} at $Ma = 0.1$ on a series of grids 
with increasing resolution.
Specifically, we simulate TGVs on $64^3$, $128^3$, $256^3$, and $384^3$ cells.
We use the two numerical setups described in the previous section.  
We use JAX version 0.2.26.
As JAX-FLUIDS can handle single- and double-precision computations, we assess the performance for both data types.
Table \ref{tab:Performance} summarizes the results.
At $384^3$ cells, only the simulation setup \textit{HLLC-float32} did not exceed the memory resources of the A6000 GPU, 
compare with Table \ref{tab:Memory}.
All results reported here are averaged over 5 independent runs.
For the \textit{HLLC-float32} setup, JAX-FLUIDS achieves a performance of around 25 $\mu s$ per time step.
This corresponds to three evaluations of the right-hand side in Equation \eqref{eq:FVD} as we use TVD-RK3 time integration.
JAX-FLUIDS, therefore, provides a strong performance taking into consideration that the code is written entirely in the high-level
language of Python/JAX.
For the \textit{ROE-float32} setup, the computation of the eigenvectors and eigenvalues increases the wall clock time roughly by an order of magnitude.
For \textit{HLLC} and \textit{ROE} schemes, we observe that the single-precision calculations are between 2.5 and 3 times faster 
than the double-precision calculations.

As GPU memory is a critical resource when working with JAX, we investigate the memory consumption of JAX-FLUIDS.
The default behavior of JAX preallocates 90\% of GPU memory in order to avoid memory fragmentation.
Therefore, to monitor the actual memory consumption, we set \mintinline{python}{XLA_PYTHON_CLIENT_PREALLOCATE="false"} to disable memory preallocation
and force JAX to allocate GPU memory as needed.
Additionally, we set \mintinline{python}{XLA_PYTHON_CLIENT_ALLOCATOR="platform"} which allows JAX to deallocate unused memory.
Note that allowing JAX to deallocate unused memory incurs a performance penalty, and we only use this setting to profile the 
memory consumption. 
Table \ref{tab:Memory} summarizes the GPU memory requirements for the aforementioned simulation setups.
We refer to the documentation of JAX \cite{jax2018github} for more details on GPU memory utilization.

\begin{table}[t!]
    \begin{center}
        \begin{tabular}{ c  c  c  c  c  c } 
        \hline
        & \multicolumn{5}{c}{Mean wall clock time per cell per time step in $10^{-9} s$}\\
        \hline
                    & $32^3$ & $64^3$ & $128^3$ & $256^3$ & $384^3$  \\ [0.5ex] 
        \hline
        HLLC - float32 & 64.41 (11.87) & 24.39 (0.06) & 25.09 (0.04) &  28.36 (0.03)  & 28.11 (0.01)  \\ 
        HLLC - float64 & 98.20 (0.49) & 93.94 (0.06) & 92.55 (0.09) &  92.76 (0.04)  & -  \\ 
        ROE - float32  & 241.71 (0.70) & 302.20 (0.32) & 304.55 (0.16) &  301.78 (0.36)  & -  \\ 
        ROE - float64  & 703.40 (1.42) & 746.78 (6.17) & 759.36 (5.23) &  760.09 (6.37)  & -  \\ 
        \hline
    \end{tabular}
    \caption{Mean wall clock time per cell per time step. 
    All computations are run on an NVIDIA RTX A6000 GPU.
    The wall clock times are averaged over five runs. 
    Numbers in brackets denote the standard deviation over the five runs.}
    \label{tab:Performance}
    \end{center}
\end{table}

\begin{table}[t!]
    \begin{center}
        \begin{tabular}{ c  c  c  c  c  c } 
        \hline
        & \multicolumn{5}{c}{Memory pressure}\\
        \hline
                    & $32^3$ & $64^3$ & $128^3$ & $256^3$ & $384^3$  \\ [0.5ex] 
        \hline
        HLLC - float32 & 295.6 (1.50) & 434.4 (1.50) & 1424.4 (1.50) & 9141.6 (1.50) & 29849.2 (2.04) \\ 
        HLLC - float64 & 353.6 (2.33) & 623.6 (2.33) & 2631.6 (2.33) & 18275.6 (2.33) & - \\ 
        ROE - float32  & 626.0 (1.79) & 818.8 (2.04) & 2255.2 (2.04) & 13546.0 (1.79) & - \\ 
        ROE - float64  & 688.0 (2.83) & 1068.4 (2.33) & 3938.0 (2.83) & 26504.4 (2.33) & - \\ 
        \hline
    \end{tabular}
    \caption{GPU Memory Pressure in megabytes (MB).}
    \label{tab:Memory}
    \end{center}
\end{table}
\section{Machine Learning in JAX-FLUIDS}
\label{sec:BackwardPass}
Having showcased that the JAX-FLUIDS solver functions very well as a modern and easy-to-use fluid dynamics simulator,
we now discuss its capabilities for ML research, in particular its automatic differentiation capabilities for end-to-end optimization.
In this section, we demonstrate that we can successfully differentiate through the entire JAX-FLUIDS solver.
We validate the AD gradients for single- and two-phase flows.
We then showcase the potential of JAX-FLUIDS by training a data-driven Riemann solver 
leveraging end-to-end optimization.

\subsection{Deep Learning Fundamentals}
\label{subsec:Deeplearning}
Given a data set of input-outputs pairs $\mathcal{D} = \left\{(\mathbf{x}_1, \mathbf{y}_1), ..., (\mathbf{x}_N, \mathbf{y}_N)\right\}$ with $\mathbf{x} \in \mathcal{X}$ and $\mathbf{y} \in \mathcal{Y}$,
supervised learning tries to find a function $f: \mathcal{X} \rightarrow \mathcal{Y}$ which (approximately) minimizes an average loss
\begin{align}
    \mathcal{L} = \frac{1}{N} \sum_{i=1}^{N} L(\mathbf{y}_i, f(\mathbf{x}_i)),
\end{align}
where $L: \mathcal{Y} \times \mathcal{Y} \rightarrow \mathbb{R}$ is a suitable loss function.
$\mathcal{X}$ and $\mathcal{Y}$ are input and output spaces, and $f \in \mathcal{F}$, where $\mathcal{F}$ is the hypothesis space.
We use $\mathbf{\hat{y}}_i$ to denote the output of the function $f$ for input $\mathbf{x}_i$, $\mathbf{\hat{y}}_i = f(\mathbf{x}_i)$.
A popular loss in regression tasks is the mean-squared error (MSE) 
\begin{align}
    \mathcal{L} = MSE(\mathbf{x}, \mathbf{y}) = \frac{1}{N} \sum_{i=1}^{N} (\mathbf{y}_i - \mathbf{\hat{y}}_i)^2.
\end{align}
One possible and highly-expressive class of functions are deep neural networks (DNN) \cite{Lecun2015,Goodfellow2016}.
DNNs are parameterizable nonlinear compound functions, $f = f_{\mathbf{\theta}}$, where the network parameters consist of weights and biases, $\mathbf{\theta} = \left\{ \mathbf{W}, \mathbf{b} \right\}$.
DNNs consist of multiple hidden layers of units between input layer and output layer.
The values in each layer are called activations.
Multilayer perceptrons (MLPs) are one particular kind of DNN in which adjacent layers are densely connected \cite{Goodfellow2016}.
We compute the activations $\mathbf{a}^l$ in layer $l$ from the activations of the previous layer $\mathbf{a}^{l-1}$,
\begin{align}
    \mathbf{a}^l = \sigma (\mathbf{W}^{l-1} \mathbf{a}^{l-1} + \mathbf{b}^{l-1}).
\end{align}
Here, $\mathbf{W}^{l-1}$ is the weight matrix linking layers $l-1$ and $l$, $\mathbf{b}^{l-1}$ is the bias vector, and $\sigma(\cdot)$
is the element-wise nonlinearity.
Typically, DNNs are trained by minimizing $\mathcal{L}$ via mini-batch gradient descent or more advanced optimization routines 
like AdaGrad \cite{Duchi2011} or Adam \cite{Kingma2015}.

\subsection{Optimization through PDE Trajectories}
Machine learning has the potential to discover and learn novel data-driven numerical algorithms for the numerical computation of fluid dynamics.
For supervised learning, we need input-ouput pairs.
In fluid dynamics, exact solutions rarely exist for complex flow.
We usually take highly-resolved numerical simulations as exact.

In the context of differentiable physics, end-to-end optimization usually refers to supervised learning of ML models 
which receive gradients that are backpropagated through a differentiable physics simulator.
Here, the ML model is placed inside a differentiable PDE solver.
A trajectory obtained from a forward simulation of the PDE solver is compared to a ground truth trajectory, and the derivatives of the loss are 
propagated across the temporal sequence, i.e., the trajectory.
We denote a trajectory of states as
\begin{align}
    \mathbf{\tau} = \left\{ \mathbf{U}^1,...,\mathbf{U}^{N_T} \right\}.
\end{align}
The differentiable solver, e.g., JAX-FLUIDS, can be interpreted as a parameterizable generator $\mathcal{G}_{\mathbf{\theta}}$ of such a trajectory 
starting from the initial condition $\mathbf{U}_0$.
\begin{align}
    \mathbf{\tau}^{PDE}_{\mathbf{\theta}} = \left\{ \mathbf{U}^1,...,\mathbf{U}^{N_T} \right\} = \mathcal{G}_{\mathbf{\theta}}(\mathbf{U}_0)
\end{align}
Our loss objective is the difference between the trajectory $\mathbf{\tau}^{PDE}_{\mathbf{\theta}}$ and a ground truth trajectory $\mathbf{\hat{\tau}} = \left\{\mathbf{\hat{U}}^1,...,\mathbf{\hat{U}}^{N_T}\right\}$.
For example, using the MSE in state space
\begin{align}
    \mathcal{L}^{\tau} = \frac{1}{N_T} \sum_{i=1}^{N_T} MSE(\mathbf{U}^i, \mathbf{\hat{U}}^i).
\end{align}
The derivatives of the loss with respect to the tuneable parameters $\partial \mathcal{L}/\partial \mathbf{\theta}$ are backpropagated 
across the simulation trajectory and through the entire differentiable PDE solver.
And the ML model is optimized by using $\partial \mathcal{L}/\partial \mathbf{\theta}$ in a gradient-based optimization routine.

In particular, multiple steps through the simulator can be chained together.
Thereby, the ML model observes the full dynamics of the underlying PDE and learns how its actions influence the entire simulation trajectory.
Naturally, the trained model is equation-specific and physics-informed.
This training procedure alleviates the problem of distribution mismatch between training and test data as the model sees its own outputs
during the training phase.
Additionally, the ML model could potentially account for approximation errors of other parts of the solver.

JAX-FLUIDS allows us to take gradients through an entire CFD simulation trajectory by applying \mintinline{python}{jax.grad} to any scalar observable of the state trajectory.
We want to stress that JAX-FLUIDS thereby differentiates for each time step through complex subfunctions such as the spatial reconstruction, 
Riemann solvers, or two-phase interactions.

\subsection{Validation of Automatic Differentiation Gradients}
\label{subsec:gradientvalid}
\begin{figure}[t]
    \centering
    \begin{tikzpicture}

    \node (A1) at (0,3.25) {};
    \node (A2) at (0,0.0) {};
    \draw (A1) -- (A2.center) node[midway, inner sep=0] (C) {};

    \draw[->] ([xshift=-0.75cm,yshift=0.25cm]A1.north) -- ([xshift=0.75cm,yshift=0.25cm]A1.north) node[midway, above] {$M_S$};

    \node[above left = 0.0cm and 0.1cm of C] (L1) {$u_L$}; 
    \node[below left = 0.0cm and 0.1cm of C] (L2) {$p_L$}; 
    \node[above = 0.0cm of L1] (L3) {$\rho_L$};
    \node[below = 0.0cm of L2] (L4) {$\gamma_L$}; 

    \node[above right = 0.0cm and 0.1cm of C] (L1) {$u_R$}; 
    \node[below right = 0.0cm and 0.1cm of C] (L2) {$p_R$}; 
    \node[above = 0.0cm of L1] (L3) {$\rho_R$};
    \node[below = 0.0cm of L2] (L4) {$\gamma_R$}; 

    \begin{axis}[
    at={(3cm,0.0cm)},
    anchor=south west,
    width=4cm, height=4cm,
    scale only axis,
    axis y line = left,
    axis x line = bottom,
    xlabel={$x$},              
    ylabel={$S$},  
    xmin=-0.1,xmax=1.1,
    ymin=0.2,ymax=1.1,
    ymajorticks=false,
    yticklabels={},
    xtick={0, 0.5, 1},
    ]

    \draw[name path=t0, line width=1] (axis cs:0, 1) -- (axis cs:0.5, 1) -- (axis cs:0.5, 0.25) -- (axis cs:1,0.25);
    \draw[name path=t1, dashed, line width=1] (axis cs:0, 1) -- (axis cs:0.75, 1) -- (axis cs:0.75, 0.25) -- (axis cs:1,0.25);
    \node at (axis cs: .625, .625) {$\Delta S$};

    \definecolor{color0}{RGB}{0,101,189}

    \addplot[color0, opacity=.4] fill between[of = t0 and t1];
    \end{axis}


    \definecolor{color0}{RGB}{0,101,189}
    \definecolor{color1}{RGB}{227,114,34}
    \definecolor{color2}{RGB}{162,173,0}

    \begin{axis}[
    anchor=south west,
    at={(9cm,0.0cm)},
    width=4cm, height=4cm,
    scale only axis,
    log basis x={10},
    log basis y={10},
    tick align=outside,
    tick pos=left,
    legend cell align={left},
    legend style={
      at={(0.03,0.97)},
      anchor=north west,
      nodes={scale=0.8, transform shape}
    },
    xlabel={\(\displaystyle \varepsilon\)},
    xmin=7.07945784384137e-05, xmax=0.141253754462275,
    xmode=log,
    xtick style={color=black},
    ylabel={\(\displaystyle \vert g_{AD} - g_{FD}^{\varepsilon} \vert_1\)},
    ymin=1.0e-8, ymax=1.0e0,
    ymode=log,
    ytick style={color=black}
    ]
    \addplot [line width=1, color0, mark=*, mark size=1, mark options={solid}]
    table {figures/fig_twophase_schematic/gradient_check_single-000.dat};
    \addlegendentry{Single-Phase}
    \addplot [line width=1, color1, mark=*, mark size=1, mark options={solid}]
    table {figures/fig_twophase_schematic/gradient_check_two-000.dat};
    \addlegendentry{Two-Phase}
    \addplot [line width=1, black]
    table {figures/fig_twophase_schematic/gradient_check_two-001.dat};
    \addlegendentry{\(\displaystyle \mathcal{O}(\Delta x^2)\)}
    \end{axis}

\end{tikzpicture}
    \caption{Left: Schematic of the computational setup. 
    The constant initial states are separated by a single right-running shock discontinuity.
    Middle: Schematic of the total entropy increase.
    Right: Error convergence of the gradient obtained by second-order central finite-difference approximation with respect to the 
    gradient obtained by automatic differentiation, i.e., $\vert g_{AD} - g_{FD}^{\varepsilon} \vert_p $.}
    \label{fig:gradients}
\end{figure}
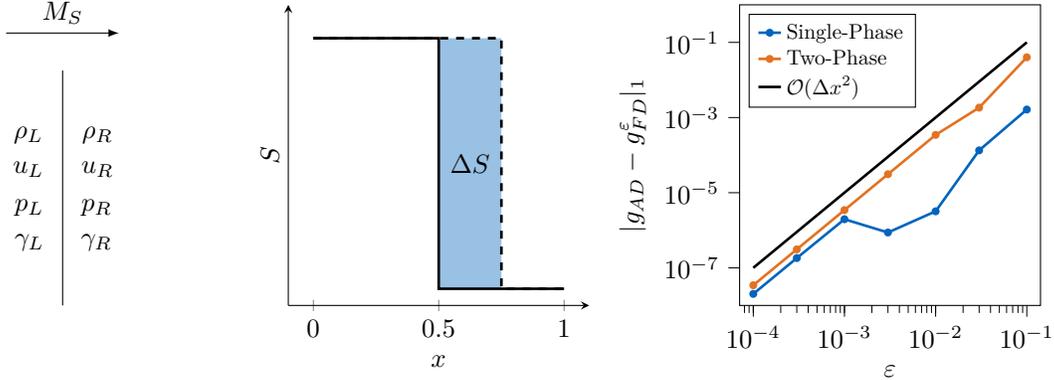

Before we showcase the full potential of JAX-FLUIDS for learning data-driven numerical schemes by end-to-end optimization,
we first validate the gradients obtained from automatic differentiation by comparing them with gradients obtained from finite-differences.
We consider a single shock wave with shock Mach number $M_S$ propagating into a fluid at rest.
We fix the state ahead of the shock with $\rho_R = p_R = 1, u_R = 0$.
The Rankine-Hugoniot relations \cite{Toro2009a} determine the state behind the shock wave as a function of the shock Mach number $M_S$.
As the shock wave crosses the computational domain, the integral entropy increases.
The described setup is depicted in Figure \ref{fig:gradients} on left.
The left and right states are separated by an initial shock discontinuity.
We consider a single-phase and two-phase setup.
In both setups all fluids are modeled by the ideal gas law.
In the former, the same fluid is left and right of the initial shock wave and $\gamma_L = \gamma_R$.
In the latter, two immiscible fluids are separated by the shock wave, i.e., $\gamma_L \neq \gamma_R$.
For the second setup, we make use of the entire level-set algorithm as described earlier.

As the shock wave propagates into the domain, the integral entropy in the domain increases by $\Delta S$,
see the schematic in the middle of Figure \ref{fig:gradients}.
The integral entropy at time $t$ is defined by 
\begin{align}
    S(t) = \int_\Omega \rho(x,t) s(x,t) dx,
\end{align}
and the increase in integral entropy is
\begin{align}
    \Delta S(t) = S(t) - S(t_0) = \int_\Omega \left(\rho(x,t) s(x,t) - \rho(x,t=0) s(x,t=0) \right) dx.
\end{align}
In the simplified setting under investigation, the increase in integral entropy at any fixed point in time, say $t^n$, is solely determined by the shock Mach number $M_S$,
i.e., $\Delta S^n = \Delta S(t=t^n) = \Delta S(M_S)$.
We let the shock wave propagate for $n$ steps with fixed $\Delta t$, i.e., $t^n = n \Delta t$, and compute the gradient
of the total entropy increase $\Delta S^n$ with respect to the shock Mach number $M_S$.
\begin{align}
    g = \frac{\partial \Delta S^n}{\partial M_S}
\end{align}
We compute the gradient with automatic differentiation $g_{AD}$ and with second-order central finite-differences according to
\begin{align}
    g_{FD}^{\varepsilon} = \frac{\Delta S^n (M_S + \varepsilon) - \Delta S^n (M_S - \varepsilon)}{2 \varepsilon}.
\end{align}
Here, we set $M_S = 2$.
We use the \textit{HLLC} setup described in the previous section.
We set $\Delta t = 1 \times 10^{-2}$ and $n = 5$.
In the single-phase case $\gamma_L = \gamma_R = 1.4$, in the two-phase case $\gamma_L = 1.4, \gamma_R = 1.667$.
On the right of Figure \ref{fig:gradients}, we visualize the $l_1$ norm between the gradients $g_{AD}$ and $g_{FD}^{\varepsilon}$ for the single-phase and two-phase setup.
We choose $\varepsilon \in \left[ 1e-1, 3e-2, 1e-2, 3e-3, 1e-3, 3e-4, 1e-4 \right]$.
We observe that the finite-difference approximations converge with second-order to the respective automatic differentiation gradients.
We conclude that automatic differentiation through the entire JAX-FLUIDS code works and gives correct gradients.
In passing, we note that although the described setting is a simplified one-dimensional setup, and we only differentiate w.r.t a single parameter,
the AD call has to backpropagate through multiple integration steps with several Runge-Kutta substeps each accompanied by calls to spatial reconstruction and Riemann solver.
Especially in the two-phase setup, gradients are obtained through the entire level-set computational routines including 
level-set advection, level-set reinitilization, and the extension procedure.  

\subsection{End-to-end Optimization of a Riemann Solver}
\label{subsec:rusanovnn}

Automatic differentiation yields the opportunity to optimize and learn numerical schemes from data by end-to-end optimization through a numerical simulator \cite{Bar-Sinai2019,Bezgin2021a,Kochkove2101784118}.
In this section, we want to showcase how JAX-FLUIDS is able to learn a numerical flux function (i.e., an approximate Riemann solver) by minimizing a loss between predicted trajectory and ground truth trajectory.
We optimize the popular Rusanov flux function (also known as local Lax-Friedrichs flux function).
The Rusanov flux at the cell face $x_{i+1/2}$ is  
\begin{align}
    \mathbf{F}_{i+1/2}^\text{Rusanov} = \frac{1}{2} (\mathbf{F}_L + \mathbf{F}_R) - \frac{1}{2} \alpha (\mathbf{U}_R - \mathbf{U}_L).
    \label{eq:Rusanov}
\end{align}
Here, $\mathbf{U}_{L/R}$ and $\mathbf{F}_{L/R}$ are the left and right sided cell face reconstructions of the conservative variable and the flux.
$\alpha$ is the scalar numerical viscosity.
For the classical Rusanov method the numerical viscosity is defined at each cell face  
$\alpha_\text{Rusanov} = \max\left\{|u_L - c_L |, |u_L + c_L|, |u_R - c_R|, |u_R + c_R|\right\}$. 
$u$ is the cell face normal velocity and $c$ is the local speed of sound.
It is well known that although the Rusanov method yields a stable solution,
the excess numerical diffusion leads to smearing out of the solution.
As a simple demonstration of the AD-capabilities of our CFD solver, we introduce the Rusanov-NN flux 
with the dissipation $\alpha^\text{NN}_\text{Rusanov} = NN \left( \vert \Delta u \vert, u_M, c_M, \vert \Delta s \vert \right)$ to be optimized.
The dissipation is output of a multi-layer perceptron which takes as inputs the jump in normal velocity $\Delta u = \vert u_R - u_L \vert$,
the mean normal velocity $u_M = \frac{1}{2} (u_L + u_R)$, the mean speed of sound $c_M = \frac{1}{2} (c_L + c_R)$, and the entropy jump $\Delta s = \vert s_R - s_L \vert$.
The network is composed of three layers with 32 nodes each.
We use RELU activations for the hidden layers.
An exponential activation function in the output layer guarantees $\alpha_\text{Rusanov}^\text{NN} \geq 0$.
\begin{figure}
    \centering
    \input{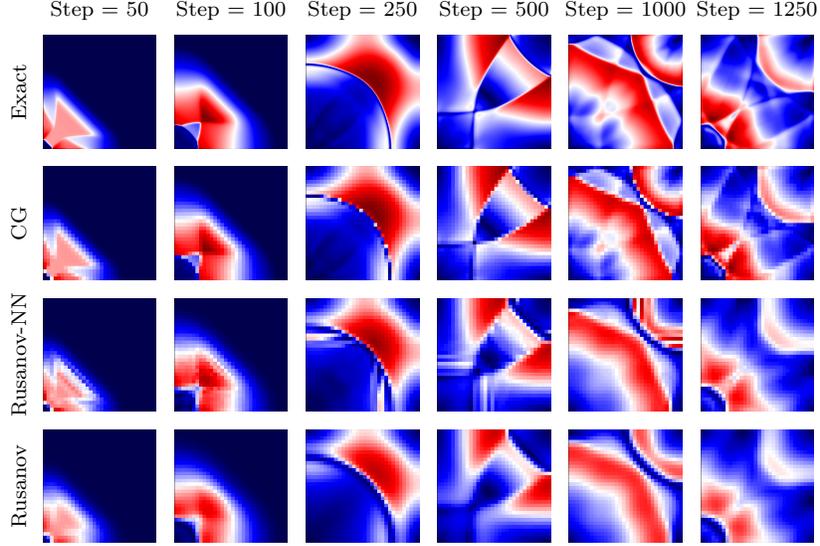}
    \caption{Trajectories of the absolute velocity. 
    From top to bottom: Ground truth (Exact) on $128 \times 128$, Coarse-grained (CG) on $32 \times 32$,
    Rusanov-NN on $32 \times 32$, and Rusanov on $32 \times 32$.
    For each time step, values are normalized with the minimum and maximum value of the exact solution.}
    \label{fig:Trajectory}
\end{figure}
We set up a highly-resolved simulation of a two-dimensional implosion test case to generate the ground truth trajectory.
The initial conditions are a diagonally placed jump in pressure and density, 
\begin{align}
    \left( \rho, u, v, p \right) = \begin{cases}
        \left(0.14, 0, 0, 0.125\right)    & \text{if}\ x + y \leq 0.15,\\
        \left(1, 0, 0, 1\right)  & \text{if}\ x + y > 0.15,
    \end{cases}
    \label{eq:FullImplosion}
\end{align}
on a domain with extent $x \times y \in [0, 1] \times [0, 1]$.
A shock, a contact discontinuity, and a rarefaction wave emanate from the initial discontinuity and travel along the diagonal.
The shock is propagating towards the lower left corner and is reflected by the walls resulting in a double Mach reflection,
while the rarefaction wave is travelling into the open domain.
The high resolution simulation is run on a mesh with $128 \times 128$ cells with a WENO3-JS cell face reconstruction, TVD-RK2 integration scheme, and the HLLC Riemann solver.
We use a fixed time step $\Delta t = 2.5 \times 10^{-4}$ and sample the trajectory every $\Delta t_\text{CG} = 10^{-3}$, 
which is the time step for the coarse grained trajectories.
A trajectory of 2501 time steps is generated, i.e., $t \in [0, 2500 \Delta t_\text{CG}]$.
Exemplary time snapshots for the absolute velocity are visualized in the top row of Figure \ref{fig:Trajectory}.
We obtain the ground truth data after coarse-graining the high-resolution trajectory onto $32 \times 32$ points, see second row of Figure \ref{fig:Trajectory}.

\begin{figure}[t!]
    \centering
    \input{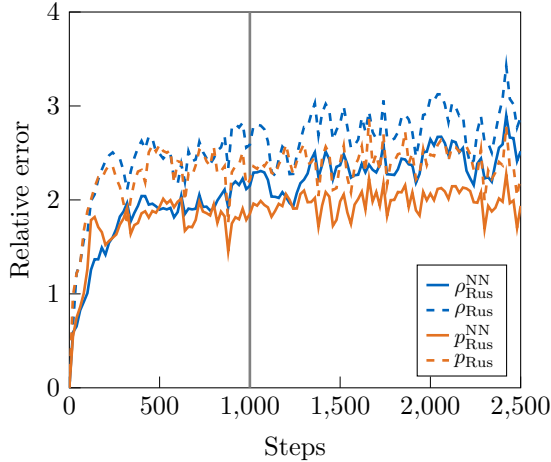}
    \caption{Relative $L_1$ error for density (blue) and pressure (orange). The gray line indicates the training horizon.}    
    \label{fig:RelError}
\end{figure}

The dissipation network model is trained in a supervised fashion by minimizing the loss between the coarse-grained (CG) trajectory and the simulation produced by the Rusanov-NN model.
The loss function is defined as the mean-squared error between the predicted and coarse-grained primitive state vectors, $\mathbf{W}^\text{NN}$ and $\mathbf{W}^\text{CG}$, over a trajectory of length $N_T$, 
\begin{align}
    L = \frac{1}{N_T}\sum_{i = 1}^{N_T} MSE(\mathbf{W}^\text{NN}_i, \mathbf{W}^\text{CG}_i).
    \label{eq:Loss}
\end{align}
The training data set consists of the first $1000$ time steps of the coarse grained reference solution. 
During training, the model is unrolled for $N_T = 15$ time steps. 
We use the Adam optimizer with a constant learning rate $5e-4$ and a batch size of $20$.
The Rusanov-NN model is trained for $200$ epochs.
Although we have trained on trajectories of length $15$, the trained model is then evaluated for the entire trajectory of $2501$ time steps. 
Figure \ref{fig:Trajectory} compares the results of the Rusanov and the final Rusanov-NN flux functions.
The NN-Rusanov flux is less dissipative than the classical Rusanov scheme and recovers small scale flow structures very well, e.g., see time step 100 in Figure \ref{fig:Trajectory}.
The NN-Rusanov flux stays stable over the course of the simulation and consistently outperforms the Rusanov flux, 
see the relative errors in pressure and density in Figure \ref{fig:RelError}.
The ML model even performs very well outside the training set (time steps larger than $1000$).
\section{Conclusion}
\label{sec:Conclusion}
We have presented JAX-FLUIDS, a comprehensive state-of-the-art fully-differentiable python package for compressible three-dimensional computational fluid dynamics.
Machine learning is becoming more and more dominant in the physical and engineering sciences.
Especially, fluid dynamics represents a field in which ML techniques and data-driven methods show very promising results and seem to have a high potential.
Despite the recent surge of ML-CFD research, a comprehensive state-of-the-art differentiable CFD solver has not been published.
JAX-FLUIDS provides researchers at the intersection of fluid dynamics and deep learning the opportunity to explore new data-driven numerical models for fluid dynamics.
JAX-FLUIDS offers powerful high-order numerical methods for a wide variety of fluid dynamics problems, e.g., turbulent flows, flows with arbitrary solid boundaries, compressible flows, and two-phase flows.
The modular architecture of JAX-FLUIDS makes integration of custom submodules easy and straightforward.

Although JAX-FLUIDS covers a wide range of flow physics, some intriguing and complex phenomena like combustion, fluid-structure interaction, or cavitation cannot yet be modeled with JAX-FLUIDS.
For the future, we plan to implement appropriate numerical methods.

Currently, by far the largest limitation of JAX-FLUIDS is the available memory of the GPU.
Although JAX-FLUIDS scales out of the box to problem sizes with roughly 400 million degrees of freedom (DOFs) on a single modern GPU,
many problems in fluid dynamics require higher DOFs.
Generally, there are two ways of tackling this problem:
adaptive multiresolution \cite{Harten1994,Harten1995} and parallelization to multiple GPUs, e.g. \cite{Romero2020,Hafner2021}. 
Adaptive multiresolution increases the mesh resolution in areas of interest (e.g., where small scale structures are present) while using a much coarser resolution in 
other parts of the flow field.
Compared to a uniform mesh, adaptive multiresolution increases the efficiency of the simulation in terms of computational resources.
To the knowledge of the authors, multiresolution strategies seem to be problematic with a jit-compiled code framework such as JAX, as the computegraph has to be static.
A second approach for increasing the computable problem size is parallelization.
The latest version of JAX \cite{jax2018github} as well as other works \cite{Hafner2021b} propose different parallelization strategies for JAX algorithms and provide promising avenues for future work.

\section*{Acknowledgements}

\appendix
\section{Numerical and Case Setup Files for Sod Shock Tube Test Case}
\label{sec:Appendix}
\begin{figure}
\begin{minted}{json}
{
    "conservatives": {
        "halo_cells": 4,
        "time_integration" : {
            "integrator": "RK3",
            "CFL": 0.9
        },
        "convective_fluxes": {
            "convective_solver": "GODUNOV",
            "riemann_solver": "HLLC",
            "signal_speed": "EINFELDT",
            "spatial_reconstructor": "WENO5-JS",
            "is_safe_reconstruction": true,
            "reconstruction_var": "PRIMITIVE"
        },
        "dissipative_fluxes": {
            "reconstruction_stencil_face": "RC4",
            "derivative_stencil_center": "DC4",
            "derivative_stencil_face": "DF4"
        }
    },
    "active_physics": {
        "is_inviscid_flux": true,
        "is_viscous_flux": false,
        "is_heat_flux": false,
        "is_volume_force": false
    },
    "active_forcings": { 
        "is_mass_flow_forcing": false,
        "is_temperature_forcing": false
    },
    "output":  {
        "is_double_precision_compute": true,
        "is_double_precision_output": true,
        "derivative_stencil": "DC4",
        "quantities": {
            "primes": ["density", "velocityX", "pressure"]
        }
    }
}
\end{minted}
    \caption{Numerical setup \textit{json} file for the Sod shock tube test case.}
    \label{fig:numerical_setup}
\end{figure}

\begin{figure}
\begin{minted}{json}
{
    "case_name": "sod",
    "nx": 100,
    "ny": 1,
    "nz": 1,
    "domain_size": {
        "x": [ 0.0, 1.0 ],
        "y": [ 0.0, 1.0 ],
        "z": [ 0.0, 1.0 ]
    },
    "end_time": 0.2,
    "save_path": "./results",
    "save_dt": 0.01,
    "boundary_types": {
        "east": "neumann",
        "west": "neumann",
        "north": "inactive",
        "south": "inactive",
        "top": "inactive",
        "bottom": "inactive"
    },
    "initial_condition": {
        "rho": "lambda x: 1.0 * (x <= 0.5) + 0.125 * (x > 0.5)",
        "u": 0.0,
        "v": 0.0,
        "w": 0.0,
        "p": "lambda x: 1.0 * (x <= 0.5) + 0.1 * (x > 0.5)"
    },
    "gravity": [ 0.0, 0.0, 0.0 ],
    "material_properties": {
        "type": "IdealGas",
        "dynamic_viscosity": 1e-4,
        "bulk_viscosity": 0.0,
        "thermal_conductivity": 0.0,
        "specific_heat_ratio": 1.4,
        "specific_gas_constant": 1.0
    },
    "nondimensionalization_parameters": {
        "density_reference": 1.0,
        "length_reference": 1.0,
        "velocity_reference": 1.0,
        "temperature_reference": 1.0
    }
}
\end{minted}
    \caption{Case setup \textit{json} file for the Sod shock tube test case.}
    \label{fig:case_setup}
\end{figure}




\section*{Data Availability Statement}
JAX-FLUIDS is available under the MIT license at \url{https://github.com/tumaer/JAXFLUIDS}.

\newpage
\bibliography{paper.bbl}

\end{document}